\documentclass[aps,prc,onecolumn,amsmath,11pt,superscriptaddress,floatfix,nofootinbib,preprintnumbers]{revtex4-2}  

\usepackage{mathrsfs} 
\usepackage{amsfonts}
\usepackage{amsmath,amssymb,bm}
\usepackage{array}
\usepackage{verbatim}
\usepackage{epsfig}
\usepackage{graphicx}
\usepackage[colorlinks=true,citecolor=green,linkcolor=blue,urlcolor=blue]{hyperref}
\usepackage[normalem]{ulem}
\usepackage{xcolor}
\usepackage{multirow}

\usepackage{graphicx}
\graphicspath{{./figure/}}
\usepackage{dcolumn}
\usepackage{bm}
\usepackage{slashed}
\usepackage{siunitx}
\usepackage{xpatch}
\usepackage{hyperref}
\usepackage[mathlines]{lineno}
\usepackage{comment}
\usepackage{soul}
\usepackage{xfrac}
\newcommand{\diff}{{\rm d}}

\usepackage{float}
\hypersetup{colorlinks, linkcolor = [rgb]{0, 0, 0.5}, citecolor = [rgb]{0,0.0,1.0}, urlcolor = [rgb]{0,0.0,0.5}}
\allowdisplaybreaks[1]

\begin{document}

\title{Dilepton production in the photodisintegration of the deuteron}

\newcommand*{\PKU}{School of Physics, Peking University, Beijing 100871, China}\affiliation{\PKU}
\newcommand*{\SDU}{Key Laboratory of Particle Physics and Particle Irradiation (MOE), Institute of Frontier and Interdisciplinary Science, Shandong University, Qingdao, Shandong 266237, China}\affiliation{\SDU}
\newcommand*{\CHEP}{Center for High Energy Physics, Peking University, Beijing 100871, China}

\author{Mengchu~Cai}\affiliation{\PKU}
\author{Tianbo~Liu}\email{liutb@sdu.edu.cn}\affiliation{\SDU}
\author{Bo-Qiang~Ma}\email{mabq@pku.edu.cn}\affiliation{\PKU}\affiliation{\CHEP}

\begin{abstract}
We study the lepton pair production in the photodisintegration of the deuteron process. The complete seven-fold differential cross section is calculated via the Bethe-Heitler mechanism with final state interactions taken into account. The deuteron bound state is described by a relativistic covariant deuteron-nucleon vertex. With numerical results, we find that the differential cross section has strong dependence on the lepton azimuthal angle in the small polar angle region and sharp peaks appear in the dependence on the invariant mass of the produced lepton pair or the two nucleons in the final state. We demonstrate that such nearly singular feature originates from the collinearity between the produced lepton or antilepton and the incident photon, and it is physically regularized by the lepton mass in our calculation. The final state interaction between the knocked-out nucleon and the recoil nucleon redistributes the differential cross section over the missing momentum, with a significant enhancement at large missing momentum and a suppression in the intermediate region. With a further decomposition of the final state interaction contribution, it is found that the on-shell term dominates the near quasi-elastic region while the off-shell term dominates the other end. In addition, we examine the contribution from the interference between the proton amplitude and the neutron amplitude, which as expected is found negligible even if the proton-neutron rescattering is included. The result in this work can serve as an input for the analysis and background estimation of multiple exclusive measurements at Jefferson Lab and future electron-ion colliders.
\end{abstract}
\maketitle

\section{Introduction}

The deuteron is the simplest nontrivial nucleus in nature, predominantly consisting of a proton-neutron bound state, and is considered an ideal place to learn the structure of cold nuclear matter and the nucleon-nucleon interaction. It has been studied from both experimental and theoretical aspects for more than 80 years since the discovery and is nowadays still an active frontier in nuclear and particle physics. Since the electromagnetic interaction is better understood, the electron and photon scatterings from the deuteron with or without disintegration are main processes to probe the deuteron structure~\cite{Bacca:2014tla,Gilman:2001yh,Garcon:2001sz,Boeglin:2015cha,Marcucci:2015rca}.
The near-threshold production of $J/\psi$ from a deuteron target received great interest in recent years because of its sensitivity to the gluonic field and the nucleon-nucleon short-range correlations (SRCs)~\cite{Wu:2013xma,Tu:2020ymk,JLab-proposal:E12-11-003B}. Furthermore, the final-state interaction in the incoherent scattering allows to investigate the $J/\psi\,N$ scattering cross section at low energies~\cite{Wu:2013xma,Tu:2020ymk,Anikin:2017fwu}. As the $J/\psi$ in the final state is usually reconstructed from its decay channels to $e^+e^-$ and $\mu^+\mu^-$, the lepton pair production is one of the dominant background reactions for the precise measurement of the $J/\psi$ production cross section, especially near the threshold.

Apart from the study of nuclear structures, the deuteron is also utilized as an effective neutron target because of the lacking of a stable free neutron target in the exploration of nucleon structures. The deep-inelastic electron scattering from the deuteron plays an important role in the flavor separation of parton distribution functions and meanwhile provides valuable information of the nuclear matter modification of the partonic structure of a nucleon~\cite{Strikman:2017koc,Cosyn:2020kwu}. Beyond the parton distribution functions, which describe the longitudinal momentum distribution of quarks and gluon in the nucleon, one of the main goals of current and future nuclear experiments is to have precise measurement of three-dimensional partonic structures of the nucleon. The generalized parton distribution (GPD), which encodes the transverse coordinate distribution of partons, is one of physical quantities to be investigated at Jefferson Lab and future electron-ion colliders~\cite{Anderle:2021wcy,AbdulKhalek:2021gbh,Accardi:2012qut}. A golden channel to access GPDs is the deeply virtual Compton scattering (DVCS)~\cite{Ji:1996nm}, in which one can practically extract GPDs from the interference term between the DVCS amplitude and the Bethe-Heitler (BH) amplitude~\cite{JeffersonLabHallA:2007jdm,Berger:2001zb,Cosyn:2018rdm,Kirchner:2003wt}, while new processes with unique sensitivities to the $x$-dependence of GPDs are recently suggested~\cite{Qiu:2022bpq,Qiu:2022pla}. Several other processes extended from the DVCS, such as the timelike Compton scattering (TCS)~\cite{Berger:2001xd,Boer:2015fwa,Boer:2015cwa,Boer:2016lcn} and the deeply virtual exclusive meson production (DVMP)~\cite{Collins:1996fb,Belitsky:2005qn,Diehl:2003ny}, are also proposed as complementary channels to measure GPDs via the similar mechanism. In such measurements, the Bethe-Heitler process is a dominant background channel, which should be well understood. 


The dilepton production in the photodisintegration of the deuteron through the BH mechanism shares the same final state particles as those in the TCS process and the $J/\psi$ production with subsequent decay to a lepton pair from the deuteron. The BH contribution dominates the total cross section over the typical TCS region by more than one order of magnitude~\cite{Berger:2001xd,Boer:2015fwa}, and is thus a main source of background preventing the precise measurement of GPDs. For the $J/\psi$ production, the cross section is small and changes rapidly when the beam energy is near the threshold, and hence the yield from the BH channel is one of the key sources of systematic uncertainties in the analysis of the $J/\psi$ production. Besides, the final state interaction (FSI) also plays an important role in certain kinematic regions where the missing momentum of the undetected nucleon is not small~\cite{CiofidegliAtti:2004jg,Laget:2004sm,JeffersonLabHallA:2004tad,Schiavilla:2005hz,Jeschonnek:2008zg}, and thus can shed light on the extraction of the $J/\psi\, N$ scattering cross section. All these above require a careful study of the BH lepton pair production from the photodisintegration of the deuteron, and it is necessary to include the FSI effects~\cite{Wu:2013xma,JLab-proposal:E12-11-003B}. 

In this paper, we perform a complete calculation of the differential cross section of the dilepton production process in the photodisintegration of the deuteron through the BH mechanism. The lepton mass is kept finite not only to differentiate the results of $e^+e-$ and $\mu^+\mu^-$ production but also to physically regularize the collinear singularities. A relativistic covariant spectator theory~\cite{Gross:1972ye,Buck:1979ff,Gross:1991pm,Gross:2010qm} is utilized to describe the deuteron in terms of the proton and the neutron, which has been proven successful in describing the elastic and inelastic electron deuteron scatterings~\cite{Jeschonnek:2008zg,VanOrden:1995eg,Adam:2002cn} and the deuteron magnetic moment and electromagnetic form factors~\cite{Gross:2014wqa,Gross:2019thk}. 
The FSI is taken into account via the nucleon-nucleon scattering, which is parametrized in a Lorentz covariant form including all spin-dependent contributions following the formalism in~\cite{Jeschonnek:2008zg}, although other approaches like the Glauber approximation or the generalized eikonal approximation are also commonly used in previous studies~\cite{Frankfurt:1996xx,Debruyne:2000wh,CiofidegliAtti:2000xj,CiofidegliAtti:2004jg,Laget:2004sm,Sargsian:2004tz,Schiavilla:2005hz}. With numerical results, we further analyze the impact of the FSI in comparison with the calculation using the plane wave impulse approximation (PWIA) and its kinematic dependence.

The remaining of this paper is organized as follows. In Sec.~\ref{theory_frame}, we present the theoretical framework of the calculation, including the kinematics of the process, the covariant deuteron-nucleon vertex, and the nucleon-nucleon rescattering in the final state. Both the PWIA and the FSI amplitudes are derived with detailed provided in appendix. In Sec.~\ref{numerical_result}, we show the numerical results for the differential cross section with detailed discussions of the nearly singular behavior in some particular kinematic regions and the effects of FSI which is further decomposed into an on-shell term and an off-shell term to understand the kinematic dependence. Summary and conclusions are given in Sec.~\ref{conclusions}.

\section{Theoretical calculation \label{theory_frame}}

\begin{figure}[h]
    \includegraphics[width=0.4\textwidth]{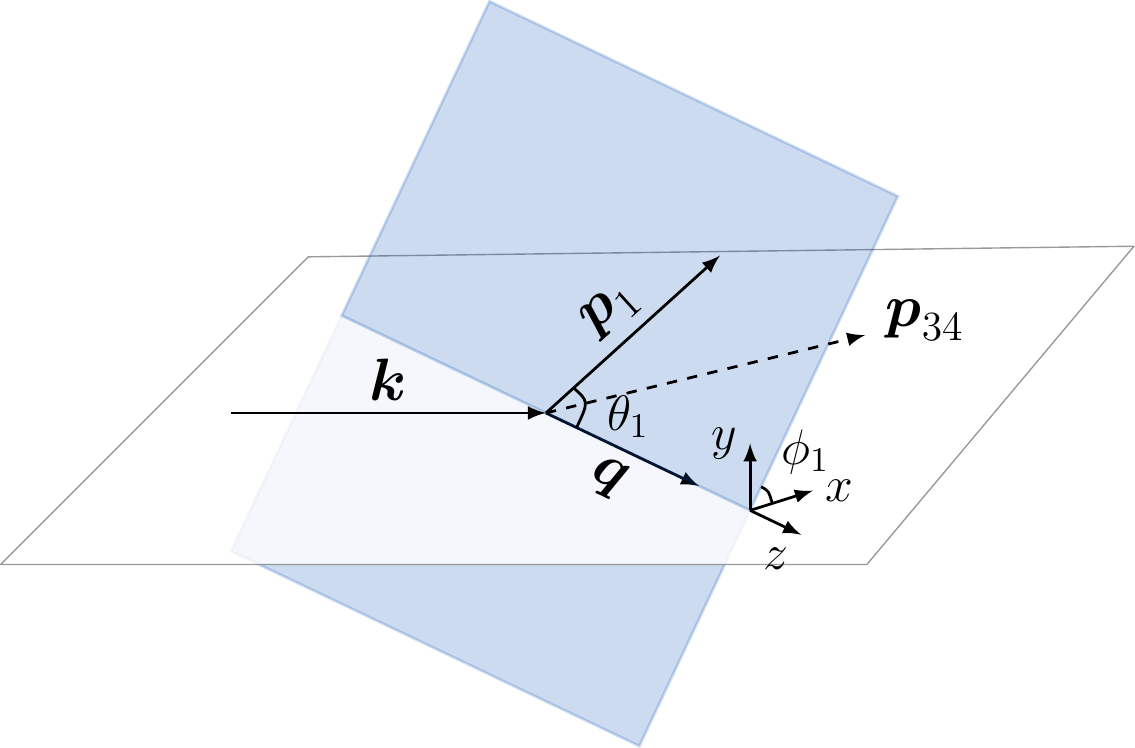}
    \hspace{3em}
    \includegraphics[width=0.4\textwidth]{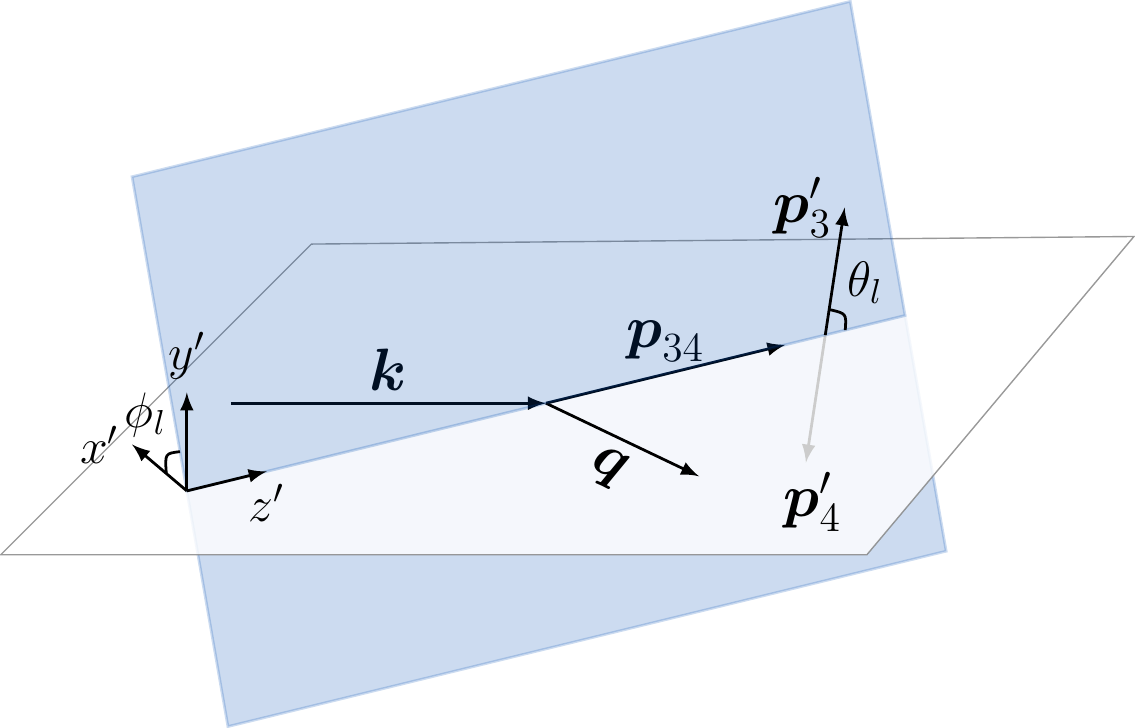}
    \caption{Kinematics for the dilepton photoproduction on the deuteron target.}
    \label{fig_kine_plane}
\end{figure}

We consider the process,
\begin{equation}
    \gamma(k) + d(P) \rightarrow p(p_1) + n(p_2) + l^-(p_3) + l^+(p_4),
\end{equation}
where the variables in parentheses represent the four-momenta of corresponding particles. The $l^-$ and $l^+$ are a pair of charged leptons, such as $e^-e^+$ or $\mu^- \mu^+$. The total four-momentum of the lepton pair in the final state is $p_{34} = p_3 + p_4$ and the four-momentum transferred to the hadronic system is $q = k - p_{34} = p_1 + p_2 - P$.
To describe the angular distribution of final state particles, it is convenient to specify a reference frame. Here we choose the deuteron rest frame, or sometimes referred to as the target rest frame, where 
\begin{align}
    P &= (m_d, {\bm 0}),
    \quad
    k = (E_\gamma, {\bm k}),
    \quad
    q = (\nu, {\bm q}).
\end{align}
As illustrated in Fig.~\ref{fig_kine_plane}, we choose the photon momentum ${\bm k}$ and the transferred momentum ${\bm q}$ in the $x-z$ plane. The direction of the proton momentum ${\bm p}_1$ is labeled by the polar angle $\theta_1$ and the azimuthal angle $\phi_1$ with respect to ${\bm q}$. The neutron momentum ${\bm p}_2$, which is not shown in Fig.~\ref{fig_kine_plane}, is fixed by the momentum conservation ${\bm p}_2 = {\bm q} - {\bm p}_1$, and we denote $|{\bm p}_2| = p_m$ as the so-called missing momentum. The direction of the final-state lepton momentum ${\bm p}_3$ (or ${\bm p}_4$) is easier to be defined in the center-of-mass frame of the lepton pair, where we label the polar angle and the azimuthal angle of $l^-$ as $\theta_l$ and $\phi_l$ with respect to the direction of ${\bm p}_{34}$ in the target rest frame.

With the kinematic variables defined above, we can write the differential cross section in the target rest frame as
\begin{eqnarray}
 \diff \sigma &=&  (2\pi)^4 \delta^{(4)}(k+P-p_1-p_2-p_3-p_4)\nonumber\\
 &\times &\frac{1}{4E_\gamma m_d}\overline{\sum_{\rm if}}|\mathcal{M}|^2 \frac{{\rm{d}^3} \bm{p}_1}{(2\pi)^3 2 E_1}  \frac{{\rm{d}^3} \bm{p}_2}{(2\pi)^3 2 E_2}  \frac{{\rm{d}^3} \bm{p}_3}{(2\pi)^3 2 E_3} \frac{{\rm{d}^3} \bm{p}_4}{(2\pi)^3 2 E_4}~,
 \label{dcs_equation}
\end{eqnarray}
where $E_i=\sqrt{m_i^2+\bm{p}_i^2}$ is the energy of corresponding particles, $m_1 = m_2 =m$ is the nucleon mass, and $m_3 = m_4 = m_l$ is the lepton mass. The symbol $\displaystyle {\overline{\sum}_{\rm if}}$ represents the average over spin states in the initial state and the sum over spin states in the final state. 

In this study, we consider the lepton pair production via the Bethe-Heitler mechanism. With one-photon-exchange approximation, we can express the invariant amplitude square as
\begin{equation}
    \overline{\sum_{\rm if}}|\mathcal{M}|^2 = \frac{e^6}{t^2}L^{\mu\nu}W_{\mu\nu}~,
    \label{ampsq}
\end{equation}
where $t=q^2$ is the transferred momentum square. The unpolarized leptonic tensor $L^{\mu\nu}$ is given by 
\begin{eqnarray}
    L^{\mu \nu}&=&-\frac{1}{2} \operatorname{Tr}\left\{\left(\slashed p_{3}+m_l\right)
    \left[\gamma^{\alpha} \frac{\slashed{p}_{3}-\slashed{k}+m_l}{\left(p_{3}-k\right)^{2}-m_l^{2}} \gamma^{\mu}\right.+\gamma^{\mu} \frac{\slashed{k}-\slashed{p}_{4}+m_l}{\left(k-p_{4}\right)^{2}-m_l^{2}} \gamma^{\alpha}\right] \nonumber\\
    & \times& \left. \left(\slashed{p}_{4}-m_l\right)
    \left[\gamma^{\nu} \frac{\slashed{p}_{3}-\slashed{k}+m_l}{\left(p_{3}-k\right)^{2}-m_l^{2}} \gamma_{\alpha} + \gamma_{\alpha} 
    \frac{\slashed{k}-\slashed{p}_{4}+m_l}{\left(k-p_{4}\right)^{2}-m_l^{2}} \gamma^{\nu}\right]\right\}~,
    \label{lepton_tensor}
\end{eqnarray}
which is the same as the one for the Bethe-Heitler process~\cite{Heller:2019dyv}. The information of nuclear structure is contained in the hadronic tensor,
\begin{equation}
    W_{\mu\nu}=(2m)^2\overline{\sum_{\rm{if}}}J_\mu(q) J_\nu^{\dagger}(q)~,
    \label{eq:Wmunu}
\end{equation}
where $J_\mu(q)=\left\langle\bm{p}_{1} s_{1} ; \bm{p}_{2} s_{2}\left|\hat{J}_{\mu}(0)\right| \bm{P} \lambda_{d}\right\rangle$ is the nuclear electromagnetic current matrix element in the momentum space, and $\lambda_d$, $s_1$, and $s_2$ represent the spin states of the initial deuteron, the final proton, and the final neutron respectively. A factor $(2m)$ is inserted for each nucleon spinor to be consistent with the normalization convention in following calculations. The current conservation requires
\begin{equation}
    q^\mu J_\mu(q) = \nu J^0(q) - \bm{q} \cdot \bm{J}(q)=0~.
\end{equation}
If taking $\bm{q}$ along the $z$-direction, one can replace the explicit $J_z(q)$ dependence by $\left(\nu/|\bm{q}|\right) J^0(q)$ in the calculation.

\begin{figure}[h]
    \includegraphics[width=0.4\textwidth]{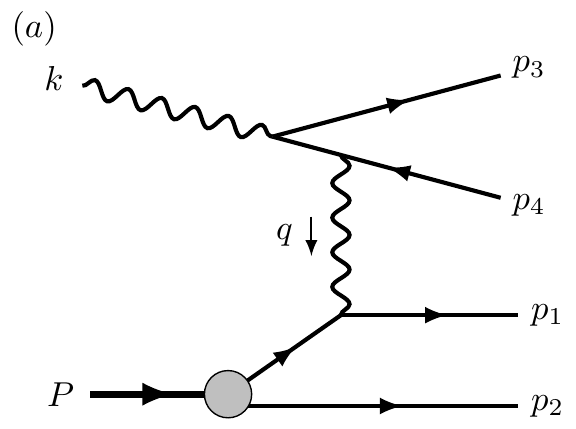}\hspace{3em}
    \includegraphics[width=0.4\textwidth]{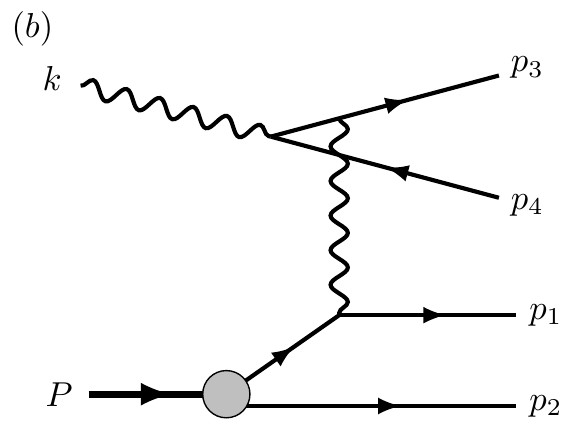}
    \caption{The Feynman diagrams for the PWIA in the one-photon-exchange approximation.}
    \label{feynpw}
\end{figure}

After integrating the four-momentum conservation $\delta$-function and a global azimuthal angle of final-state particles with respect to the photon axis, we obtain a seven-fold differential cross section. For convenience, we adopt the invariant mass square of the proton and the neutron $s_{pn}=(p_1+p_2)^2$ and the invariant mass square of the lepton pair $s_{ll}=(p_3+p_4)^2$ as variables. Then the differential cross section can be expressed as
\begin{equation}
    \diff \sigma =\frac{\alpha^3}{8(4\pi)^4}\frac{|\bm{p}_1|^2\beta}{m_d^2 E_\gamma^2 t^2}\frac{L^{\mu\nu}W_{\mu\nu}}{||\bm{p}_1|(m_d+\nu)-E_1|\bm{q}|\cos \theta_1|}{\rm{d}} \Omega_p {\rm{d}} \Omega_{ll}{\rm{d}}s_{ll}{\rm{d}}s_{pn}{\rm{d}}t~,
    \label{dcs_tot}
\end{equation}
where $\Omega_p$ is the solid angle of the proton with $\diff\Omega_p = \diff\cos\theta_1 \diff\phi_1$, $\Omega_{ll}$ is the solid angle of $l^-$ in the lepton pair center-of-mass frame with $\diff\Omega_{ll} = \diff\cos\theta_l \diff\phi_l$, $\beta$ is the lepton velocity given by
\begin{equation}
    \beta=\sqrt{1-\frac{4m_l^2}{s_{ll}}}~,
\end{equation}
and $\alpha = e^2/(4\pi)$ is the electromagnetic fine structure constant. The derivation from Eq.~\eqref{dcs_equation} to Eq.~\eqref{dcs_tot} is provided in Appendix~\ref{dcs_formula}.

\subsection{Plane wave contribution}

Firstly, we consider the impulse approximation (IA), in which the reaction is viewed as the Bethe-Heitler process from a quasi-free nucleon in the deuteron. The mechanism of this plane wave (PW) contribution is depicted by the Feynman amplitudes in Fig.~\ref{feynpw}. In this case, electromagnetic current matrix element can be written as 
\begin{eqnarray}
    J_{\rm{PW}}^\mu(q) &=& \left\langle\bm{p}_{1} s_{1} ; \bm{p}_{2} s_{2}\left|\hat{J}^{\mu}\right| \bm{P} \lambda_{d}\right\rangle_{\rm{PW}}\nonumber\\ 
    &=&\bar{u}\left(\bm{p}_{1}, s_{1}\right) \Gamma_N^{\mu}(q) G_{0}\left(P-p_{2}\right) \Gamma_{\lambda_{d}}\left(p_{2}, P\right)C \bar{u}^{T}\left(\bm{p}_{2}, s_{2}\right)~,
    \label{pw_contri}
\end{eqnarray}
where the Dirac spinors are normalized as $u^{\dagger}({\bm p},s) u({\bm p},s') = E/m$, $C=-i\gamma^0\gamma^2$ is the charge conjugation operator, and $G_0$ is the bare nucleon propagator given by 
\begin{equation}
    G_0(p)=\frac{\slashed{p}+m}{m^2-p^2-i\epsilon}~,
    \label{bare_prop}
\end{equation}
where $\epsilon$ is a positive infinitesimal number. The nucleon electromagnetic current vertex $\Gamma_N^\mu (q)$ is parametrized as
\begin{equation}
    \Gamma_N^\mu(q)=F_1(Q^2)\gamma^\mu+\frac{F_2(Q^2)}{2m}i\sigma^{\mu\nu} q_\nu~,
    \label{N_vertex}
\end{equation}
where $Q^2 = -t$ and $F_1(Q^2)$ and $F_2(Q^2)$ are nucleon's Dirac and Pauli form factors. In numerical calculations, we use the fit of nucleon electromagnetic form factors in Ref.~\cite{Ye:2017gyb}. The deuteron-proton-neutron vertex can be written as
\begin{align}
	\Gamma_{\lambda_d}(p_2,P) = \Gamma_D^\mu(p_2,P)~\xi^{\lambda_d}_{\mu}(P),
\end{align}
where $\xi_\mu^{\lambda_d}(P)$ is the deuteron polarization vector and $\Gamma_D^\mu(p_2,P)$ is the covariant vertex function with $P$ and $p_2$ on-shell. According to the Lorentz structure, one can in general express the vertex function as~\cite{Gross:1972ye,Gross:1991pm,Gross:2010qm}
\begin{equation}
    \Gamma_{D}^\mu(k_2,P)\equiv F \gamma^\mu+\frac{G}{m} k^\mu-\frac{m-\slashed{k}_1}{m}\left(H \gamma^\mu+\frac{I}{m} k^\mu\right)~,
    \label{D_vertex}
\end{equation}
where $k = (k_2 - k_1)/2$ is the relative four-momentum of the two nucleons with $k_1 = P - k_2$ being the momentum of the off-shell nucleon. The scalar functions $F$, $G$, $H$, and $I$ encode the nucleonic structure of the deuteron and can be evaluated from the deuteron wave functions. Explicit expressions of these functions are provided in Appendix~\ref{rela_form_wave} and more details can be found in Refs.~\cite{Buck:1979ff,Gross:2014wqa}. 

Substituting~\eqref{pw_contri} into~\eqref{eq:Wmunu}, we obtain the hadronic tensor in the plane wave impulse approximation (PWIA) as
\begin{align}
W^{\mu\nu}_{\rm PW}&=
(2m)^2\overline{\sum_{\rm{if}}}J_{\rm{PW}}^\mu(q) J^{\nu\,\dagger}_{\rm{PW}}(q)
\nonumber\\
    &={\rm{Tr}}\left[\Gamma_N^{\mu}(q) \frac{\slashed{P}-\slashed{p}_2+m}{m^2-(P-p_2)^2} \Gamma_D^\alpha\left(p_{2}, P\right)\left(\slashed{p}_2-m\right)\widetilde{\Gamma}_D^\beta\frac{\slashed{P}-\slashed{p}_2+m}{m^2-(P-p_2)^2}\widetilde{\Gamma}_N^\nu \left(\slashed{p}_1+m\right)\right]\nonumber\\
	&\quad\times 
	\frac{1}{3}\sum_{\lambda_d=1}^3 \xi_\alpha^{\lambda_d}(P)~\xi^{*\,\lambda_d}_\beta(P)~,
	\label{pw_trace}
\end{align}
where
\begin{align}
    \widetilde{\Gamma}_N^\nu &=\gamma^0 \Gamma_N^{\nu\,\dagger} \gamma^0=F_1(Q^2)\gamma^\nu-\frac{F_2(Q^2)}{2m}i\sigma^{\nu\eta} q_\eta~,\\ 
    \widetilde{\Gamma}^\beta_{D}&=\gamma^0 \Gamma^{\beta\,\dagger}_{D} \gamma^0=F \gamma^\beta+\frac{G}{m} p^\beta-\left(H \gamma^\beta+\frac{I}{m} p^\beta\right)\frac{m-\slashed{p}_1}{m}~,
\end{align}
and $p=\dfrac{1}{2}(p_2-p_1+q)$. The derivation of Eq.~\eqref{pw_trace} is provided in Appendix~\ref{trace_formu}.

\subsection{Final state interactions}

\begin{figure}[h]
    \includegraphics[width=0.4\textwidth]{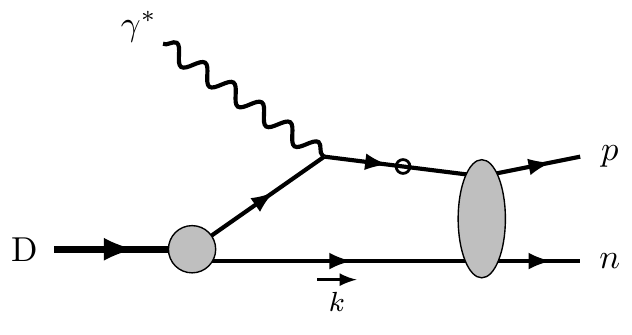}
    \caption{Final state interactions contribution to the electromagnetic current matrix element. The fermion line with ``$\circ$'' represents the particle can be off-shell and the propagator will be decomposed into positive energy contribution and negative energy contribution.}
    \label{fsi_contr}
\end{figure}

In the PWIA, the spectator nucleon does not participate in the reaction. Although this approximation is widely utilized in many Monte Carlo simulations, its validity should be examined quantitatively for precise measurement. We consider the situation that one of the nucleons is struck by the virtual photon but interacts with the other nucleon before flying out. The mechanism of such final-state interaction (FSI) is illustrated in Fig.~\ref{fsi_contr}. In this case, the proton leg and the neutron leg of the deuteron-proton-neutron vertex can be both off-shell and such kind of vertex function is explored in Ref.~\cite{Gross:2014wqa}, in which the amplitude requires a four-dimensional integration over the internal momentum $k$. In this study, we simplify the calculation by applying the covariant spectator theory~ \cite{Gross:1972ye, Buck:1979ff, Gross:1991pm, Gross:2010qm}, in which the nucleon not directly struck by the virtual photon is set on-shell and the four-dimensional integration is replaced by
\begin{equation}
    \int \frac{\diff^3 \bm{k}}{(2\pi)^3}\frac{m}{E_{\bm{k}}}~.
    \label{int_k}
\end{equation}
This method has been utilized in the study of the elastic electron-deuteron scattering~\cite{VanOrden:1995eg} and the electrodisintegration of the deuteron~\cite{Jeschonnek:2008zg}, and the results are in good agreement with experimental data.

To evaluate the FSI effect, we need to calculate the rescattering between the knocked-out nucleon and the recoil nucleon. According to the Dirac matrix structure, the nucleon-nucleon (NN) scattering matrix can be parametrized as~\cite{LaFrance1980,Gross:2008ps,Jeschonnek:2008zg}
\begin{eqnarray}
    M&=F_S(s,t) \bm{1}_1 \bm{1}_2 + F_V(s,t) (\gamma^\mu)_1 (\gamma_\mu)_2 + F_T(s,t) (\sigma^{\mu\nu})_1(\sigma_{\mu\nu})_2\nonumber\\
    &+ F_P(s,t) (\gamma^5)_1 (\gamma^5)_2 + F_A(s,t) (\gamma^5 \gamma^\mu)_1 (\gamma^5 \gamma_\mu)_2~,
    \label{cov_para}
\end{eqnarray}
where the subscripts ``$1$'' and ``$2$'' of the $\gamma$-matrices indicate the spinor space of the corresponding nucleon. $F_S$, $F_V$, $F_T$, $F_P$, and $F_A$ are scalar functions of $s$ and $t$ which are the Mandelstam variables for the NN elastic scattering. Considering the spin states of the nucleons, one can alternatively expand the NN scattering matrix as~\cite{Bystricky:1976jr,LaFrance1980}
\begin{eqnarray}
    M\left(\bm{k}_{f}, \bm{k}_{i}\right)&=\frac{1}{2}\left[(a+b)+(a-b)\left(\bm{\sigma}_{1}, \bm{n}\right)\left(\bm{\sigma}_{2}, \bm{n}\right)+(c+d)\left(\bm{\sigma}_{1}, \bm{m}\right)\left(\bm{\sigma}_{2}, \bm{m}\right)\right.\nonumber\\
	&+(c-d)\left.\left(\bm{\sigma}_{1}, \bm{l}\right)\left(\bm{\sigma}_{2}, \bm{l}\right)+e\left(\bm{\sigma}_{1}+\bm{\sigma}_{2}, \bm{n}\right)\right]~,
	\label{pauli_para}
\end{eqnarray}
where ${\bm k}_i$ and ${\bm k}_f$ are momenta of the nucleon before and after the scattering in the NN center-of-mass frame, in which the basis vectors can be constructed as
\begin{equation}
    \bm{l}=\frac{\bm{k}_{f}+\bm{k}_{i}}{\left|\bm{k}_{f}+\bm{k}_{i}\right|}, \quad \bm{m}=\frac{\bm{k}_{f}-\bm{k}_{i}}{\left|\bm{k}_{f}-\bm{k}_{i}\right|}, \quad \bm{n}=\frac{\bm{k}_{i} \times \bm{k}_{f}}{\left|\bm{k}_{i} \times \bm{k}_{f}\right|}~.
\end{equation}
The $(\bm{\sigma}_j,\bm{v})$ represents the projection of the spin Pauli matrix $\bm{\sigma}$ of the $j$th particle on the direction $\bm{v}={\bm l}$, ${\bm m}$, and ${\bm n}$. The amplitudes $a$, $b$, $c$, $d$, and $e$ are complex functions of the momentum $|\bm{k}_i|=|{\bm k}_f| = \sqrt{s-4m^2}/2$ and the scattering angle $\theta$, which can be expressed in terms of Mandelstem variables as
\begin{align}
	\cos \theta=1 + \frac{2 t}{s-4m^2}~.
\end{align}
These two parametrizations of the NN scattering matrix can be related through the helicity amplitudes as demonstrated in Appendix~\ref{para_M}. The partial wave expansion for the amplitudes $a$, $b$, $c$, $d$, and $e$ is presented in Ref.~\cite{Bystricky1987}. In numerical calculations, we use the phase shift analysis in Refs.~\cite{Arndt:1982ep,Workman:2016ysf} and the phase shift data are available from the SAID program~\cite{said_program}. 

Though the recoil nucleon is set on-shell in the covariant spectator theory, the struck nucleon can be off-shell before the rescattering. One may in principle include additional terms in the NN scattering matrix with one of the nucleon off-shell~\cite{Ford:2013zca}. In this study, we follow the approach in Ref.~\cite{Jeschonnek:2008zg} by introducing a form factor $F_N(p^2)$ to account for the off-shell effect and keep the NN scattering matrix parametrization the same form as Eq.~\eqref{cov_para}. The scalar functions $F_j(s,t)$ with $j=S$, $V$, $T$, $P$, and $A$ are then replaced by 
\begin{equation}
	F_j(s,t) \rightarrow F_j(s,t,u) F_N(s+t+u-3m^2)~,
\end{equation}
where 
\begin{equation}
    F_N(p^2)=\dfrac{(\Lambda_N^2-m^2)^2}{(p^2-m^2)^2+(\Lambda_N^2-m^2)^2}~,
    \label{off-factor}
\end{equation} 
and $p$ represents the four-momentum of the off-shell nucleon. 
In this case, the scattering angle $\theta$ is given by
\begin{align}
	\cos \theta=\frac{t-u}{\sqrt{1-\frac{4m^2}{s}}\sqrt{(4m^2-t-u)^2-4m^2 s}}~.
\end{align}
When $p^2=m^2$, the form factor $F_N(m^2)=1$ and the scattering angle $\theta$ reduces back to the regular expression for the on-shell NN scattering. $\Lambda_N$ is a phenomenological parameter, which we choose to be $1.2\,\rm{GeV}$ as suggested in Ref.~\cite{Jeschonnek:2008zg}.

\begin{figure}[h]
    \includegraphics[width=0.4\textwidth]{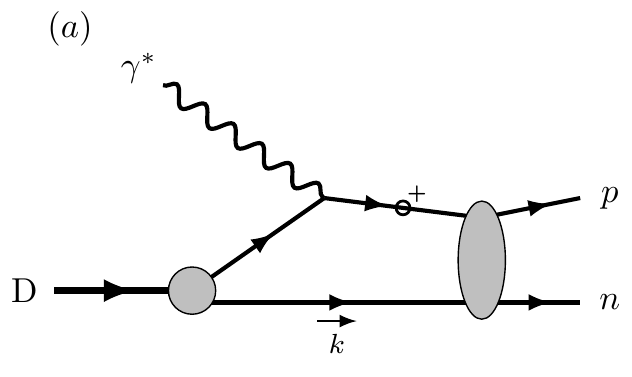}\hspace{3em}
    \includegraphics[width=0.4\textwidth]{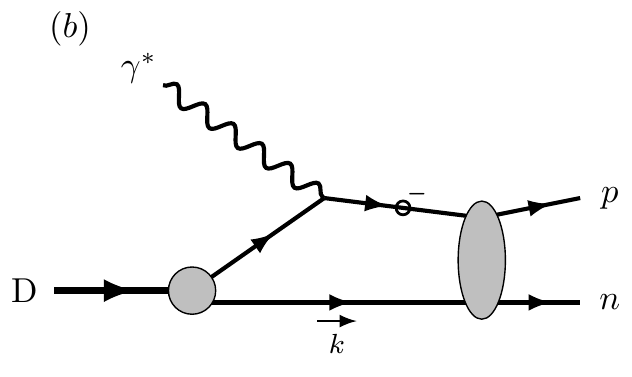}
    \caption{The decomposition of the FSI amplitude. The fermion line with ``$\circ ^{+}$'' represents the positive energy contribution, including the on-shell term and the off-shell principal value term, and the fermion line with ``$\circ ^{-}$'' represents the negative energy contribution.}
    \label{feyn_posi_nega}
\end{figure}

With Eqs.~(\ref{int_k}) and (\ref{cov_para}), the FSI contribution to the current matrix element is given by
\begin{eqnarray}
    J_{\rm{FSI}}^\mu(q)_{\rm{if}}&=&\left\langle\bm{p}_{1} s_{1}; \bm{p}_{2} s_{2}\left|\hat{J}^{\mu}\right| \bm{P} \lambda_{d}\right\rangle_{\rm{FSI}} \nonumber\\
	&=&\int \frac{d^{3} \bm{k}}{(2 \pi)^{3}} \frac{m}{E_{k}} \bar{u}_{\alpha}\left(\bm{p}_{1}, s_{1}\right) \bar{u}_{\beta}\left(\bm{p}_{2}, s_{2}\right) M_{\alpha \alpha^\prime ; \beta \beta^\prime}\left(p_{1}, p_{2} ; k\right) \nonumber\\
	& \times & G_{0\, \alpha^\prime  \eta}\left(P+q-k\right) \Gamma_{N\,\eta \eta^\prime}^{\mu}(q) G_{0\, \eta^\prime \zeta} \left(P-k\right)( \Gamma_{\lambda_{d }}C)_{\zeta \zeta^\prime}\left(k, P\right)\Lambda_{\beta^\prime \zeta^\prime}^{+}\left(\bm{k}\right)~.
	\label{fsi_integ}
\end{eqnarray}
Then we decompose the bare nucleon propagator into positive and negative energy terms,
\begin{eqnarray}
    G_{0}(p) &=&-\frac{m}{E_{p}} \sum_{s}\left[\frac{u(\boldsymbol{p}, s) \bar{u}(\boldsymbol{p}, s)}{p^{0}-E_{p}+i \epsilon}+\frac{v(-\boldsymbol{p}, s) \bar{v}(-\boldsymbol{p}, s)}{p^{0}+E_{p}-i \epsilon}\right]\nonumber \\
	&=&-\frac{m}{E_{p}}\left[\frac{\Lambda_{+}(\boldsymbol{p})}{p^{0}-E_{p}+i \epsilon}-\frac{\Lambda_{-}(-\boldsymbol{p})}{p^{0}+E_{p}-i \epsilon}\right],
\end{eqnarray}
where the projection operators $\Lambda_{\pm}$ are defined as
\begin{align}
     \Lambda_+(\bm{p})&=\frac{\slashed{\bar{p}}+m}{2m} = \sum_s u(\bm{p},s)\bar{u}(\bm{p},s)~,\\
     \Lambda_-(\bm{p})&=-\frac{\slashed{\bar{p}}-m}{2m} = -\sum_s v(\bm{p},s)\bar{v}(\bm{p},s)~,
\end{align}
corresponding to the positive and negative energy solutions respectively, and $\bar{p}$ denotes the on-shell four-momentum. 
Furthermore, we use the relation
\begin{equation}
    \frac{1}{p^{0}-E_{p}+i \epsilon}=-i \pi \delta\left(p^{0}-E_{p}\right)+\frac{\mathcal{P.V.}}{p^{0}-E_{p}}
\end{equation}
to write the positive energy term into two parts. Hence, Eq.~(\ref{fsi_integ}) is decomposed into three parts,
\begin{eqnarray}
    J_{\rm{FSI}}^\mu(q)_{\rm{if}}=J_{\rm{FSI}}^\mu(q)^{\rm{A}}_{\rm{if}}+J_{\rm{FSI}}^\mu(q)^{\rm{B}}_{\rm{if}}+J_{\rm{FSI}}^\mu(q)^{\rm{C}}_{\rm{if}}~,
\end{eqnarray}
where 
\begin{eqnarray}
    J_{\rm{FSI}}^\mu(q)^{\rm{A}}_{\rm{if}} &=& \sum_{s}\int \frac{d^{3} \bm{k}}{(2 \pi)^{3}} (i\pi)\frac{m}{E_{k}}\frac{m}{E_{P+q-k}}\delta(P^0+q^0-E_k-E_{P+q-k})\nonumber\\
    & \times &\bar{u}_{\alpha}\left(\bm{p}_{1}, s_{1}\right) \bar{u}_{\beta}\left(\bm{p}_{2}, s_{2}\right) M_{\alpha \alpha^\prime ; \beta \beta^\prime}\left(p_{1}, p_{2} ; k\right) u_{\beta^\prime}\left(\bm{k}, s\right)\nonumber\\
	& \times & \Lambda^+_{ \alpha^\prime  \eta}\left(\bm{P}+\bm{q}-\bm{k}\right) \Gamma_{N\,\eta \eta^\prime}^{\mu}(q) G_{0\, \eta^\prime \zeta} \left(P-k\right)( \Gamma_{\lambda_{d }}C)_{\zeta \zeta^\prime}\left(k, P\right)\bar{u}^T_{\zeta^\prime}\left(\bm{k}, s\right)~
	\label{fsi_onshell}
\end{eqnarray}
is the on-shell positive energy contribution,
\begin{eqnarray}
    J_{\rm{FSI}}^\mu(q)^{\rm{B}}_{\rm{if}} &=& \sum_{s}\int \frac{d^{3} \bm{k}}{(2 \pi)^{3}} (-1)\frac{m}{E_{k}}\frac{m}{E_{P+q-k}}\frac{\mathcal{P.V.}}{P^0+q^0-E_k-E_{P+q-k}}\nonumber\\
    & \times &\bar{u}_{\alpha}\left(\bm{p}_{1}, s_{1}\right) \bar{u}_{\beta}\left(\bm{p}_{2}, s_{2}\right) M_{\alpha \alpha^\prime ; \beta \beta^\prime}\left(p_{1}, p_{2} ; k\right) u_{\beta^\prime}\left(\bm{k}, s\right)\nonumber\\
	& \times & \Lambda^+_{ \alpha^\prime  \eta}\left(\bm{P}+\bm{q}-\bm{k}\right) \Gamma_{N\,\eta \eta^\prime}^{\mu}(q) G_{0\, \eta^\prime \zeta} \left(P-k\right)( \Gamma_{\lambda_{d }}C)_{\zeta \zeta^\prime}\left(k, P\right)\bar{u}^T_{\zeta^\prime}\left(\bm{k}, s\right)~
	\label{fsi_pv}
\end{eqnarray}
is the principal value integral (off-shell) positive energy contribution,
and 
\begin{eqnarray}
    J_{\rm{FSI}}^\mu(q)^{\rm{C}}_{\rm{if}} &=& \sum_{s}\int \frac{d^{3} \bm{k}}{(2 \pi)^{3}} \frac{m}{E_{k}}\frac{m}{E_{P+q-k}}\frac{1}{P^0+q^0-E_k+E_{P+q-k}-i\epsilon}\nonumber\\
    & \times &\bar{u}_{\alpha}\left(\bm{p}_{1}, s_{1}\right) \bar{u}_{\beta}\left(\bm{p}_{2}, s_{2}\right) M_{\alpha \alpha^\prime ; \beta \beta^\prime}\left(p_{1}, p_{2} ; k\right) u_{\beta^\prime}\left(\bm{k}, s\right)\nonumber\\
	& \times & \Lambda^-_{ \alpha^\prime  \eta}\left(\bm{k}-\bm{P}-\bm{q}\right) \Gamma_{N\,\eta \eta^\prime}^{\mu}(q) G_{0\, \eta^\prime \zeta} \left(P-k\right)( \Gamma_{\lambda_{d }}C)_{\zeta \zeta^\prime}\left(k, P\right)\bar{u}^T_{\zeta^\prime}\left(\bm{k}, s\right)~
	\label{fsi_nega}
\end{eqnarray}
is the negative energy contribution. 
$J_{\rm{FSI}}^\mu(q)^{\rm{A}}_{\rm{if}} + J_{\rm{FSI}}^\mu(q)^{\rm{B}}_{\rm{if}}$ is shown in Fig.~\ref{feyn_posi_nega}a and $J_{\rm{FSI}}^\mu(q)^{\rm{C}}_{\rm{if}}$ is shown in Fig.~\ref{feyn_posi_nega}b. Since the negative energy part (anti-nucleon propagation) is much smaller than the positive energy part (nucleon propagation), we ignore the contribution from Eq.~(\ref{fsi_nega}) in the calculation. This is the same approximation as adopted in Ref.~\cite{Jeschonnek:2008zg}. 

\section{Numerical results and discussions}
\label{numerical_result}

In this section, we present the numerical results for the unpolarized differential cross section. We first discuss the angular distribution of the differential cross section in the PWIA and the singular feature of the cross section at some kinematics for the $e^+e^-$ production. Then we discuss the FSI effect on the differential cross section and the impact respectively from the on-shell and the off-shell contributions. Besides, we consider the effect arising from possible interference between the proton and the neutron amplitudes.

\subsection{Angular distribution}

Since the lepton pair production via the BH process is a dominant background channel for the measurement of $J/\psi$ production near the threshold, we choose the photon energy at $8.5\,\rm GeV$, which is just above the $J/\psi$ production threshold from free nucleons and is accessible at Jefferson Lab. With four particles in the final state, the complete differential cross section is seven-fold after eliminating an azimuthal symmetric angle for the scattering from unpolarized deuteron target.

\begin{figure}[htp]
 \centering
 \includegraphics[width=0.45\textwidth]{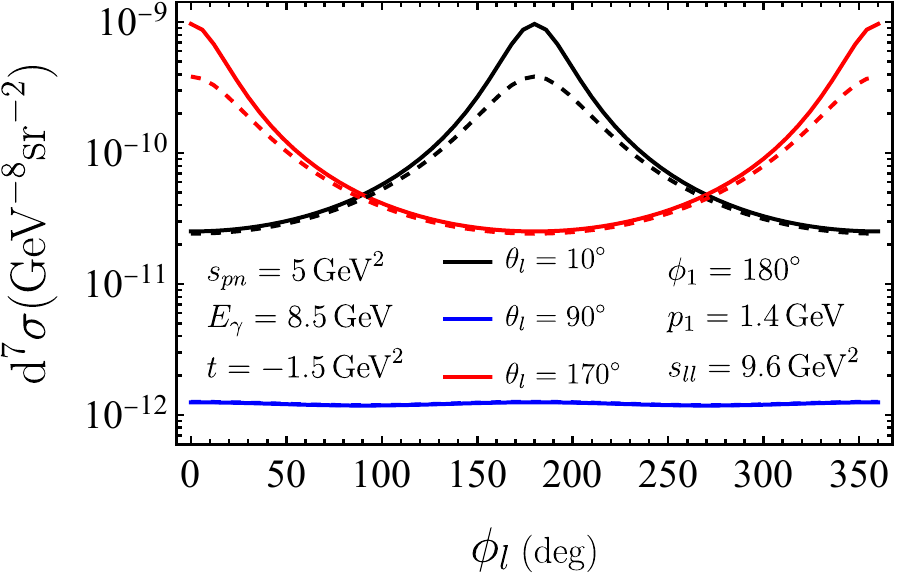}
 \caption{The differential cross sections of $e^+e^-$ production (solid curves) and $\mu^+\mu^-$ production (dashed curves) as a function of $\phi_l$ at $E_\gamma=8.5~\rm{GeV}$, $s_{pn}=5~\rm{GeV}^2$, $s_{ll}=9.6~\rm{GeV}^2$, $t=-1.5~\rm{GeV}^2$, $p_1=1.4~\rm{GeV}$, $\phi_1=180^\circ$, and $\theta_l=10^\circ,\,90^\circ,\,170^\circ$, respectively. The label $d^7 \sigma$ is a short-handed notation representing the seven-fold differential cross section ${\rm d}^7\sigma / {\rm{d}} \Omega_p {\rm{d}} \Omega_{ll}{\rm{d}}s_{ll}{\rm{d}}s_{pn}{\rm{d}}t$. The same notation is used in following figures.}
 \label{angular_phie}
\end{figure}

In Fig.~\ref{angular_phie}, we plot the differential cross section as a function of the lepton azimuthal angle $\phi_l$ at several values of $\theta_l = 10^\circ$, $90^\circ$, and $170^\circ$ to show the polar angle dependence. $s_{ll} = 9.6\,\rm GeV^2$ is set as the mass square of $J/\psi$, $s_{pn} = 5\,\rm GeV^2$ and $t = -1.5\,\rm GeV^2$ are chosen at some typical values, and $p_1 = 1.4\,\rm GeV$ and $\phi_1 = 180^\circ$ are fixed for the struck proton. As one can observe from the Fig. 5, the cross section is large when $\theta_l$ is close to $0^\circ$ or $180^\circ$. At such limit, the emitted lepton or antilepton is collinear to ${\bm p}_{34}$, the lepton pair momentum. Although the cross section does not have singularities in this kinematics region, the polar angle of ${\bm p}_{34}$ is small for the particular kinematic choice here. Hence, the enhancement around $\phi_l = 180^\circ$ or $0^\circ$ is actually from the singularity when one of the lepton pair is collinear with the incident photon as will be discussed in the next subsection. Since the electron is much lighter than the muon, the solid curves ($e^+e^-$) show sharper peaks than the dashed curves ($\mu^+\mu^-$). When the polar angle $\theta_l$ is at intermediate values, {\it e.g.} $90^\circ$ as shown in the figure, the propagator is always away from the pole and consequently leads to a smooth distribution in $\phi_l$. In such regions, the lepton mass effect is marginal and the cross sections for $\mu^+\mu^-$ and $e^+e^-$ productions are nearly the same.

\begin{figure}[htp]
 \centering
 \includegraphics[width=0.45\textwidth]{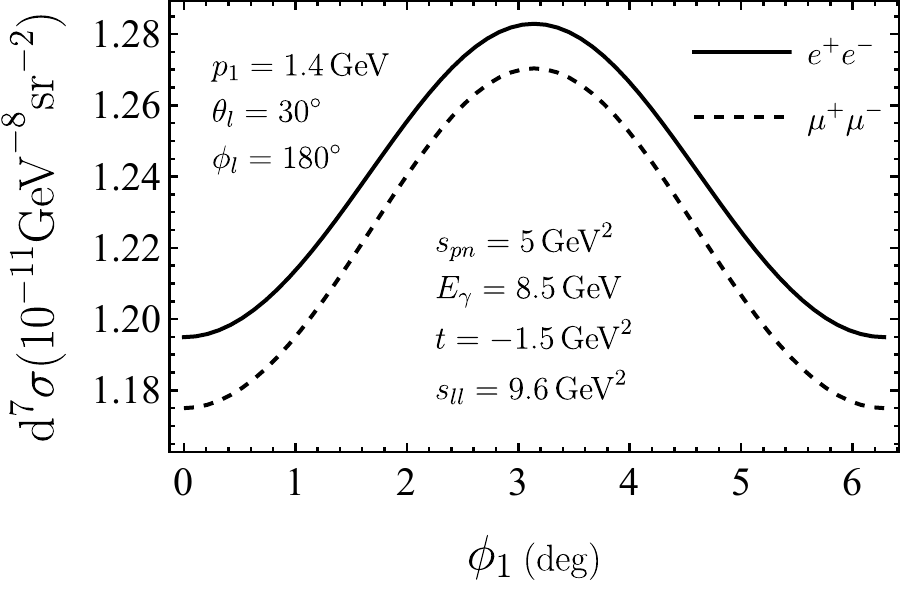}
 \caption{The differential cross sections of $e^+e^-$ production (solid curve) and $\mu^+\mu^-$ production (dashed curve) as a function of $\phi_1$ at $p_1=1.4~\rm{GeV}$, $\theta_l=30^\circ$, $\phi_l=180^\circ$, and the same $E_\gamma$, $s_{pn}$, $s_{ll}$, and $t$ values as those in Fig.~\ref{angular_phie}.}
 \label{angular_phip}
\end{figure}
 
In Fig.~\ref{angular_phip}, we plot the differential cross section as a function of $\phi_1$, the azimuthal angle of the proton. The $s_{ll}$, $s_{pn}$, and $t$ are set at the same values as Fig.~\ref{angular_phie}, and $\theta_l=30^\circ$ and $\phi_l = 180^\circ$ are fixed for the lepton angle. Since the proton acquires the large momentum from the virtual photon, which has opposite transverse momentum to the lepton pair, the struck proton tends to be around $\phi_1 = 180^\circ$. In PWIA, the smearing is caused by the fermi motion, while FSI will further broaden the distribution.

\subsection{Singular behavior for $e^+e^-$ production \label{singular_behavior}}

For massless leptons, the differential cross section has singularities if the lepton or the antilepton is collinear to the incident photon. This behavior is convenient to be understood from the BH amplitude. When the lepton is collinear to the photon, the lepton propagator in Fig.~\ref{feynpw}a approaches the pole. Similarly, the propagator in Fig.~\ref{feynpw}b reaches the pole when the antilepton is collinear to the photon. Such singularities are physically regularized by the lepton mass, which is kept nonzero in our calculation. However, a sharp peak is still observed for the $e^+e^-$ production because of the tiny mass of the electron in comparison with typical scales of the reaction presented here.

According to the physical origin of the singular behavior, it is straightforward to examine the differential cross section as a function of the lepton polar angle with respect to the incident photon. Whereas, it is not a commonly used variable in describing the lepton pair production in deuteron photodisintegration process. One has to track the singular behavior from the dependence on other kinematic variables.

\begin{figure}[htp]
\centering
\includegraphics[width=0.45\textwidth]{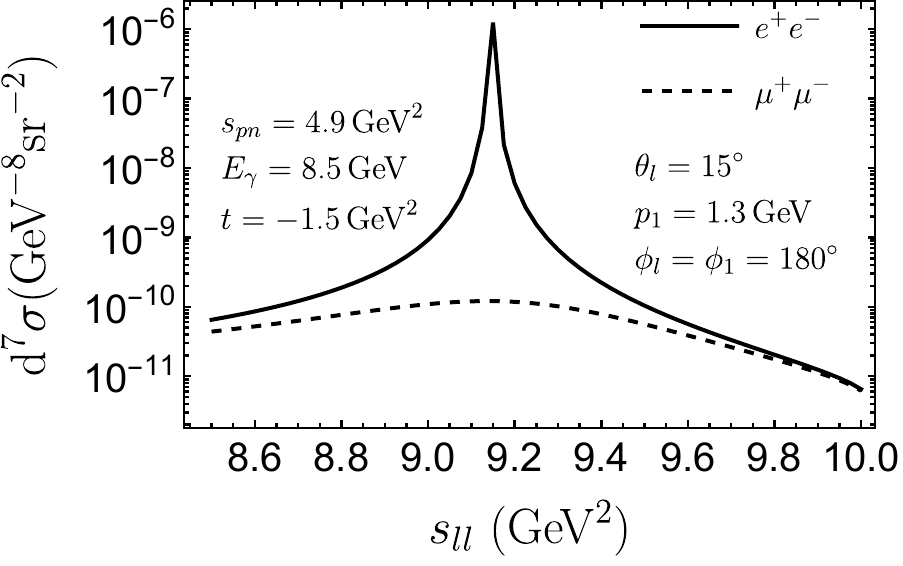}
\caption{Differential cross sections for $e^+e^-$ production (solid line) and $\mu^+\mu^-$ production (dashed line) as a function of $s_{ll}$ at $E_\gamma=8.5~\rm{GeV}$, $s_{pn}=4.9~\rm{GeV}^2$, $t=-1.5~\rm{GeV}^2$, $p_1=1.3~\rm{GeV}$,  $\theta_l=15^\circ$ and $\phi_l=\phi_1=180^\circ$.}
\label{singular_sll}
\end{figure}

\begin{figure}[htp]
   \centering
    \includegraphics[width=0.45\textwidth]{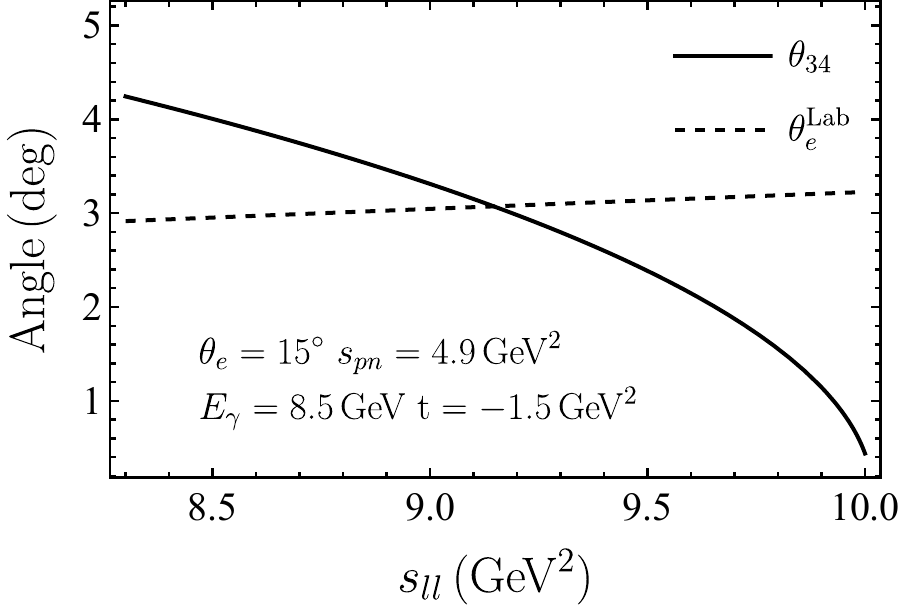}
    \caption{$\theta_{34}$ (solid line) and $\theta_e^{\rm{Lab}}$ (dashed line) vary with $s_{ll}$ at $E_\gamma=8.5~\rm{GeV}$, $s_{pn}=4.9~\rm{GeV}^2$, $t=-1.5~\rm{GeV}^2$.}
    \label{fig_theta34}
\end{figure}

In Fig.~\ref{singular_sll}, we plot the differential cross section as a function of $s_{ll}$, with other kinematic variables set at $s_{pn} = 4.9\,\rm GeV^2$, $t = -1.5\,\rm GeV^2$, $p_1 = 1.3\,\rm GeV$, $\phi_1 = 180^\circ$, $\theta_l = 15^\circ$, and $\phi_l = 180^\circ$.  One can observe that the cross section for $e^+e^-$ production sharply peaks around $s_{ll}=9.15\,\rm{GeV}^2$, where the electron is collinear to the incident photon. To clearly see this correspondence, we first examine the polar angle $\theta_{34}$ of the lepton pair momentum $\bm{p}_{34}$. At fixed $s_{pn}$ and $t$, $\theta_{34}$ can be expressed as a function of $s_{ll}$. Their relation at the chosen kinematics is drawn in Fig.~\ref{fig_theta34}. Then we consider the direction of the electron in the lab frame. The lepton-photon collinear configuration happens when $\phi_e = 180^\circ$ and $\theta_e^{\rm Lab} = \theta_{34}$, where $\theta_e^{\rm Lab}$ is the relative angle between the electron momentum ${\bm p}_3$ and the lepton pair momentum ${\bm p}_{34}$ in the laboratory frame. From the relation between $\theta_e^{\rm Lab}$ and $s_{ll}$ drawn in Fig.~\ref{fig_theta34}, one can find the intersecting point, which indicates $\theta_e^{\rm Lab} = \theta_{34}$. The corresponding $s_{ll}$ value is just at $9.15\,\rm GeV^2$. According to the definition of angles in Fig.~\ref{fig_kine_plane}, the positron-photon collinear configuration will happen at $\phi_l = 0^\circ$ and can be analyzed via the same procedure. Since muon is much heavier, the cross section of $\mu^+\mu^-$ production does not show a sharp peak.

\begin{figure}[htp]
\centering
\includegraphics[width=0.45\textwidth]{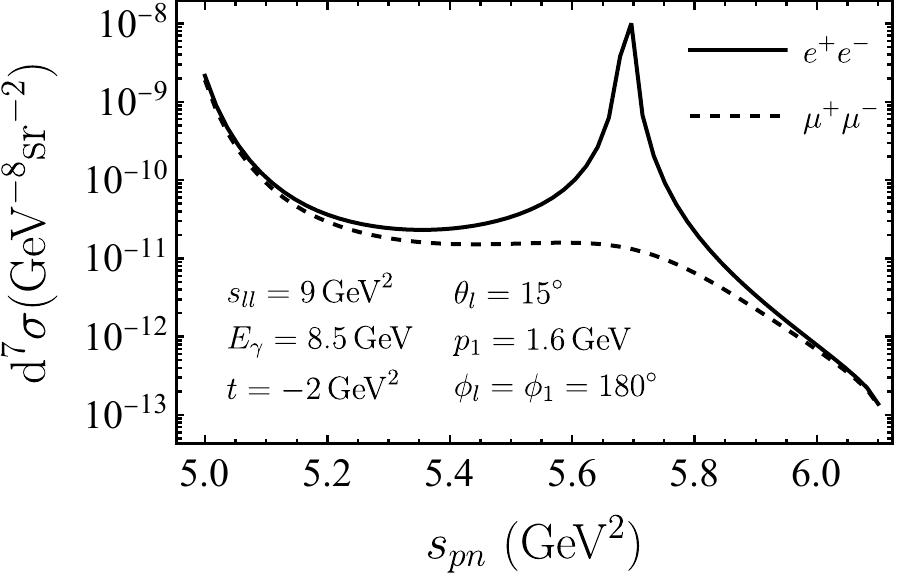}
\caption{Differential cross sections for $e^+e^-$ production (dashed line) and $\mu^+\mu^-$ production (dashed line) as a function of $s_{pn}$ at $E_\gamma=8.5~\rm{GeV}$, $s_{ll}=9~\rm{GeV^2}$, $t=-2~\rm{GeV}^2$, $p_1=1.6~\rm{GeV}$,  $\theta_l=15^\circ$ and $\phi_l=\phi_1=180^\circ$.}
\label{singular_spn}
\end{figure}

Similarly, one can also observe the sharp peak behavior in the dependence on other variables. In Fig.~\ref{singular_spn}, we plot the differential cross section as a function of $s_{pn}$, the invariant mass of the final proton-neutron system, while the other variables are set at $s_{ll}=9\,\rm{GeV^2}$, $t=-2\,\rm{GeV}^2$, $p_1=1.6\,\rm{GeV}$, $\theta_l=15^\circ$, $\phi_l=180^\circ$, and $\phi_1=180^\circ$. The cross section for $e^+e^-$ productions peaks around $s_{pn} = 5.7\,\rm GeV^2$.

We note that the peak position in $s_{ll}$ or $s_{pn}$ shifts to smaller values with increasing $\theta_l$. When $\theta_l$ is large enough, the momenta of the lepton and antilepton are always away direction of the incident photon, and thus the differential cross section has smooth behavior as a function of $s_{pn}$ or $s_{ll}$ as shown in Figs.~\ref{fig_fsi_spn} and~\ref{fig_fsi_sll}.

\subsection{Final state interactions effect}

In this subsection, we discuss the FSI contribution to the differential cross section. Because the FSI effects on the $\mu^+\mu^-$ and $e^+e^-$ productions are similar, we only present the results for the $e^+e^-$ production.

\begin{figure}[htp]
\includegraphics[width=0.45\textwidth]{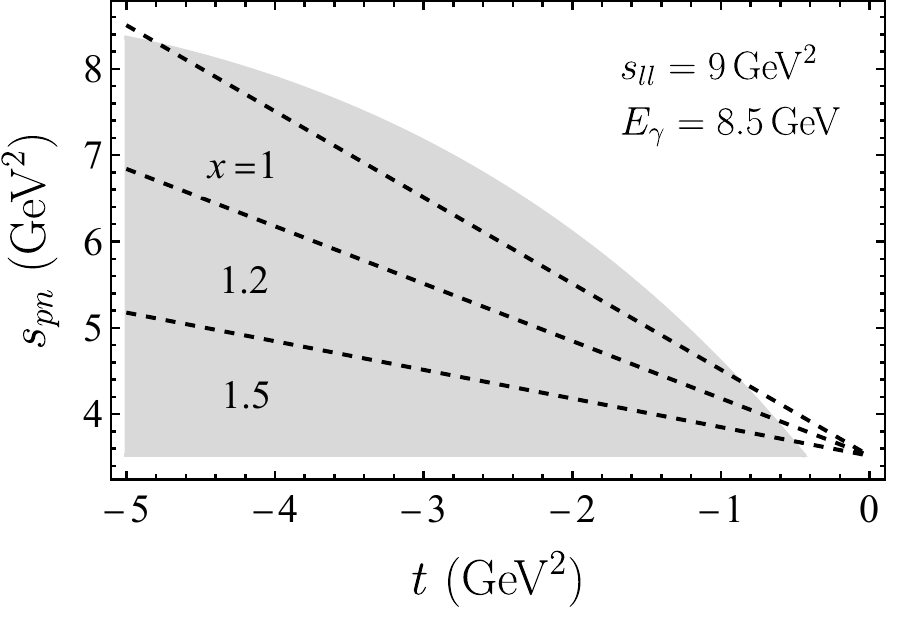}
\caption{The $s_{pn}-t$ phase space region at $E_\gamma=8.5~\rm{GeV}$ and $s_{ll}=9~\rm{GeV}^2$. The allowed kinematic range for $t$ is larger than that plotted in the figure. Dashed lines show the Bjorken $x$ at several fixed values.}
\label{fig_phasespace}
\end{figure}

In the PWIA, the reaction is approximated as the scattering from a quasi-free nucleon. As a general feature, the differential cross section decreases with increasing $|t|$. To explore the deviation from the quasi-free picture, it is convenient to introduce the Bjorken variable $x$, which goes to $1$ for elastic scattering from a free nucleon. In Fig.~\ref{fig_phasespace}, we show the phase space in the $s_{pn}-t$ plane together with lines indicating corresponding $x$ values. With fixed $s_{pn}$, the $x$ becomes closer to $1$ as $|t|$ decreases, which means closer to the quasi-free kinematics. 

\begin{figure}[htp]
\centering
\includegraphics[width=0.45\textwidth]{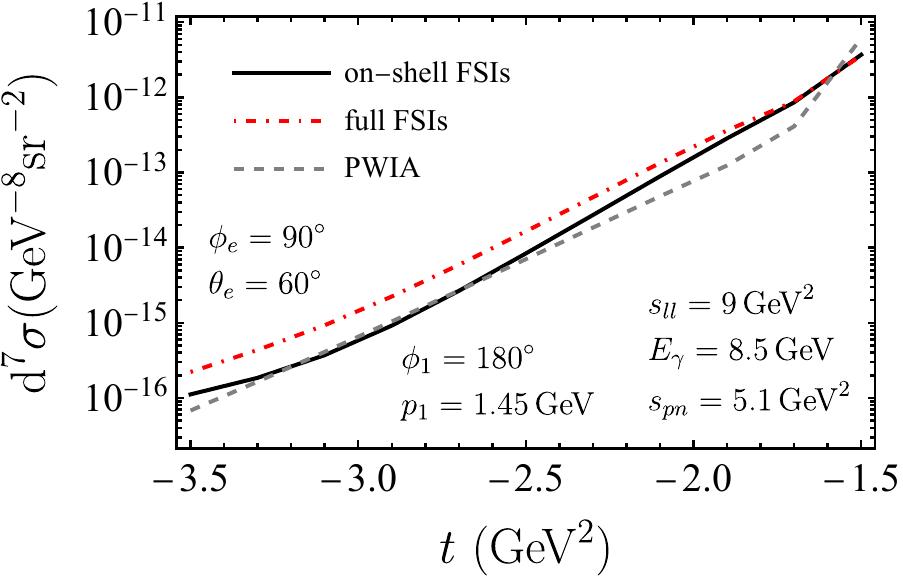}
\caption{Differential cross sections for $e^+e^-$ production as a function of $t$ at $E_\gamma=8.5~\rm{GeV}$, $s_{pn}=5.1~\rm{GeV}^2$, $s_{ll}=9~\rm{GeV}^2$, $p_1=1.45~\rm{GeV}$, $\theta_e=60^\circ$, $\phi_e=90^\circ$, and $\phi_1=180^\circ$ in three cases: in the PWIA (gray dashed curve), including the on-shell FSI (black solid curve), and including both the on-shell and the off-shell FSIs (red dot-dashed curve).}
\label{fig_fsi_t}
\end{figure}

In Fig.~\ref{fig_fsi_t}, we plot the differential cross section as a function of $t$ with other variables fixed at $s_{pn}=5.1\,\rm{GeV}^2$,  $s_{ll}=9\,\rm{GeV}^2$, $p_1=1.45\,\rm{GeV}$, $\theta_e=60^\circ$, $\phi_e=90^\circ$, and $\phi_1=180^\circ$. Instead of showing the cross section in the full $t$ region, we only present the result in $-3.5\,{\rm GeV} < t < -1.5\,{\rm GeV}$, which is viewed as a reasonable range for the approach dealing with the FSI in the current calculation. When $|t|$ is very small, {\it e.g.} $|t| < 1\,\rm GeV^2$, one may need to take into account the contribution from meson exchange current~\cite{Sargsian:2001ax}. For very large $|t|$, {\it e.g.} $|t| > 5\,\rm GeV^2$, the color coherence phenomena will complicate the situation~\cite{Sargsian:2001ax,Frankfurt:1994hf,Frankfurt:1994kt}. 
The dashed curve represents the result from PWIA, the solid curve represents the result including the on-shell FSI contribution, and the dot-dashed curve represents the result including both the on-shell and the off-shell FSI contributions.
As can be observed in Fig.~\ref{fig_fsi_t}, the full FSI curve and the on-shell FSI curve have little difference when $|t|$ is relatively small. This can be understood from Fig.~\ref{fig_phasespace}, where one may find that the $x$ becomes closer to $1$ as $|t|$ decreases with fixed $s_{pn}$. At this limit, the kinematics is close to the quasi-free scattering and thus the off-shell FSI has little impact on the cross section.

\begin{figure}[ht]
\centering
\includegraphics[width=0.45\textwidth]{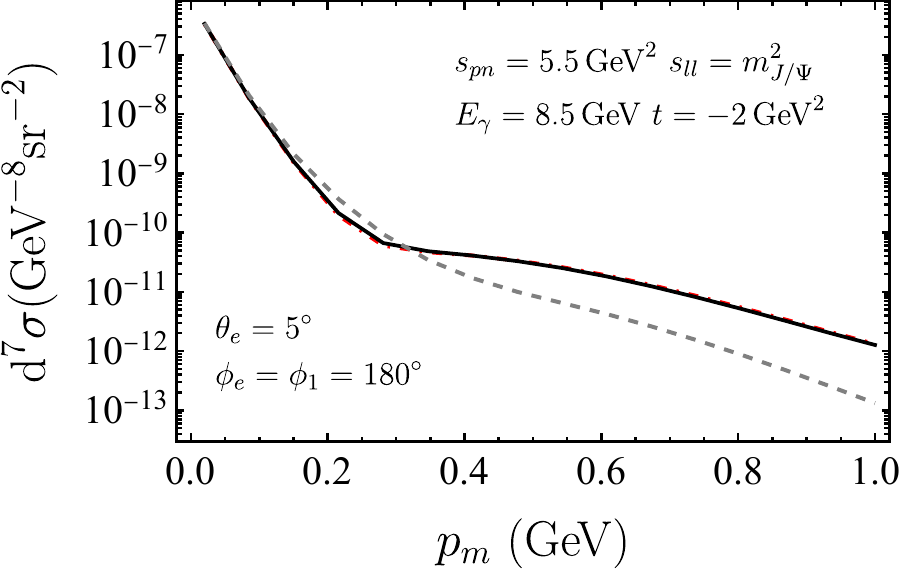}
\caption{Differential cross sections for $e^+e^-$ production as a function of the missing momentum $p_m$ at $E_\gamma=8.5~\rm{GeV}$, $s_{pn}=5.5~\rm{GeV}^2$, $t=-2~\rm{GeV}^2$, $s_{ll}=m_{J/\psi}^2$, $\theta_e=5^\circ$, and $\phi_e=\phi_1=180^\circ$. The gray dashed curve represents the result with PWIA, the black solid curve represents the result including the on-shell FSI, and the red dot-dashed curve represents the result including both the on-shell and the off-shell FSIs.}
\label{fig_fsi_pm}
\end{figure}

In Fig.~\ref{fig_fsi_pm}, we plot the differential cross section as a function of the missing momentum $p_m$, which is the amplitude of the momentum of the undetected nucleon. In the PWIA, the proton is knocked out by the virtual photon and acquires large momentum, while the spectator neutron carries relatively low momentum. Once the FSI is taken into account, the undetected neutron may received momentum transfer via the NN scattering. Therefore, one may expect that the FSI effect is more significant in the region of relatively large $p_m$. This intuitive picture is confirmed by the numerical results in Fig.~\ref{fig_fsi_pm}. The three curves are almost indistinguishable at small $p_m$, {\it e.g.} when $p_m < 0.1\,\rm GeV$. While in the large $p_m$ region, {\it e.g.} $p_m > 0.4\,\rm GeV$, significant enhancement is caused by the FSI effects. It is also interesting to find that in the medium $p_m$ region, $0.1\,{\rm GeV} < p_m < 0.3\,\rm GeV$, the cross section results including the FSI are less than the PWIA result, in which the suppression is from the interference between the PWIA amplitude and the FSI amplitude. Besides, one may find that the full FSI curve and the on-shell FSI curve have little difference in Fig.~\ref{fig_fsi_pm}. It is due to the particular kinematics, which gives $x$ close to 1, {\it i.e.} the quasi-free region, where the off-shell FSI contribution is considerably small as discussed above.

\begin{figure}[ht]
\includegraphics[width=0.45\textwidth]{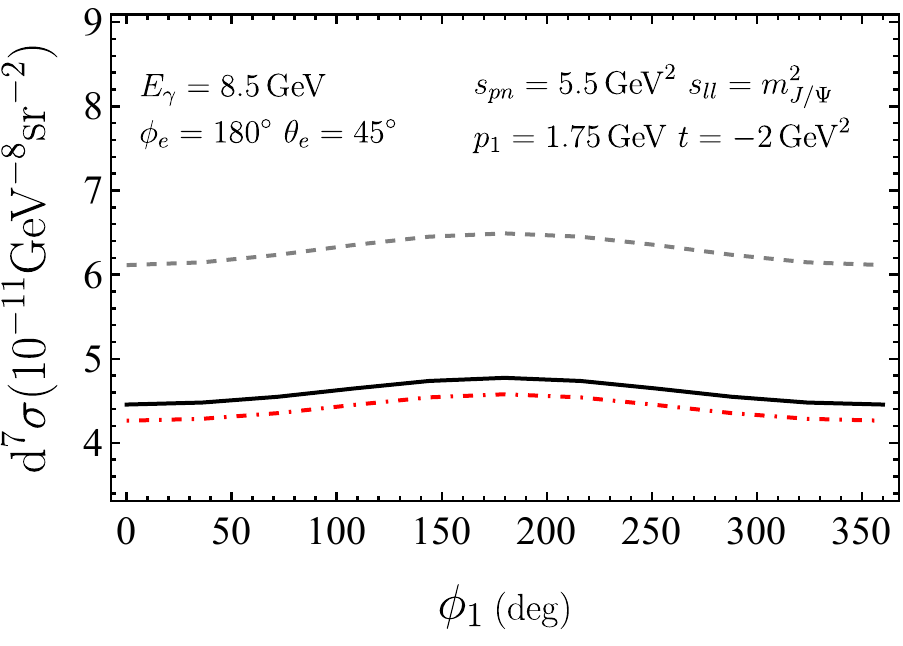}
\caption{Differential cross sections for $e^+e^-$ production as a function of $\phi_1$ at $E_\gamma=8.5~\rm{GeV}$, $s_{pn}=5.5~\rm{GeV}^2$, $t=-2~\rm{GeV}^2$, $s_{ll}=m_{J/\psi}^2$, $p_1=1.75~ \rm{GeV}$, $\phi_e=180^\circ$, and $\theta_e=45^\circ$. The gray dashed curve represents the result with PWIA, the black solid curve represents the result including the on-shell FSI, and the red dot-dashed curve represents the result including both the on-shell and the off-shell FSIs.}
\label{fig_fsi_phi1}
\end{figure}

In Fig.~\ref{fig_fsi_phi1}, we plot the differential cross section as a function of $\phi_1$ with other variables fixed at $s_{pn}=5.5\,\rm{GeV}^2$, $t=-2\,{\rm GeV}^2$, $s_{ll}=m_{J/\psi}^2$, $p_1=1.75\,\rm{GeV}$, $\phi_e=180^\circ$, and $\theta_e=45^\circ$. The final-state proton momentum $p_1=1.75~\rm{GeV}$ corresponds to the missing momentum $p_m = 0.152\,\rm{GeV}$, which is in the region where the FSI effects suppress the cross section. The on-shell FSI curve is slightly above the full FSI curve in the whole $\phi_1$ range.

\begin{figure}[ht]
    \centering
    \includegraphics[width=0.45\textwidth]{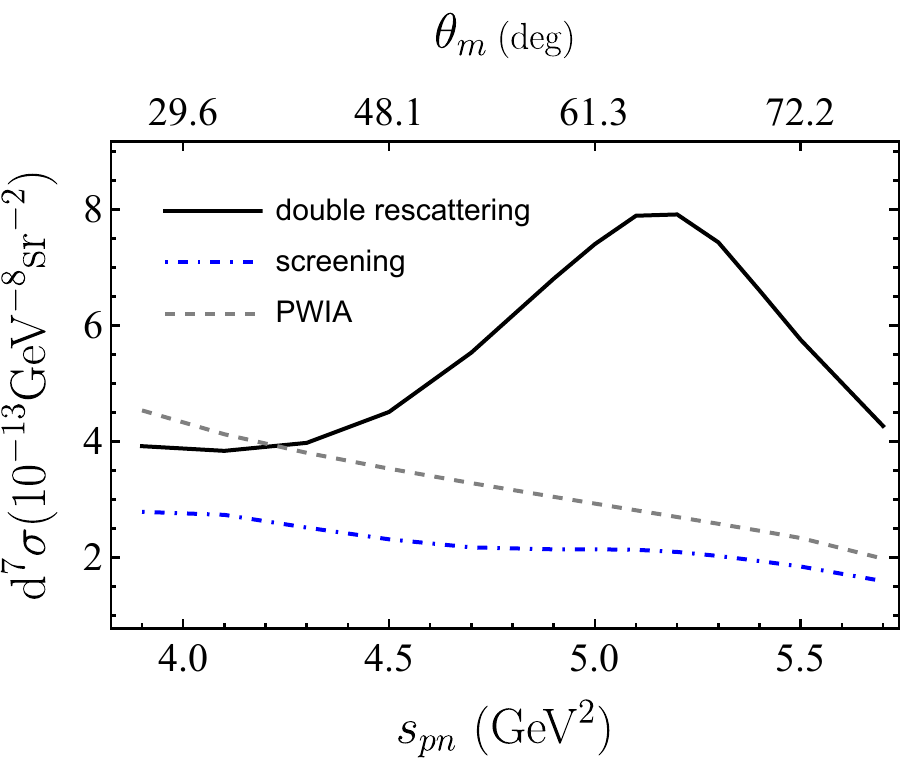}
    \caption{Differential cross sections for $e^+e^-$ production as a function of $s_{pn}$ at $E_\gamma=8.5~\rm{GeV}$, $t=-1.7~\rm{GeV}^2$, $s_{ll}=9~\rm{GeV}^2$, $p_m=0.45~\rm{GeV}$, $\phi_1=180^\circ$, $\theta_e=60^\circ$, and $\phi_e=90^\circ$ in the PWIA (gray dashed curve), including the interference between the PWIA and the FSI amplitudes (blue dot-dashed curve), and including both the interference term and the FSI amplitude square (black solid curve). The $\theta_m$ axis is nonlinear. }
    \label{fig_fsi_spn}
\end{figure}
    
\begin{figure}[ht]
    \includegraphics[width=0.45\textwidth]{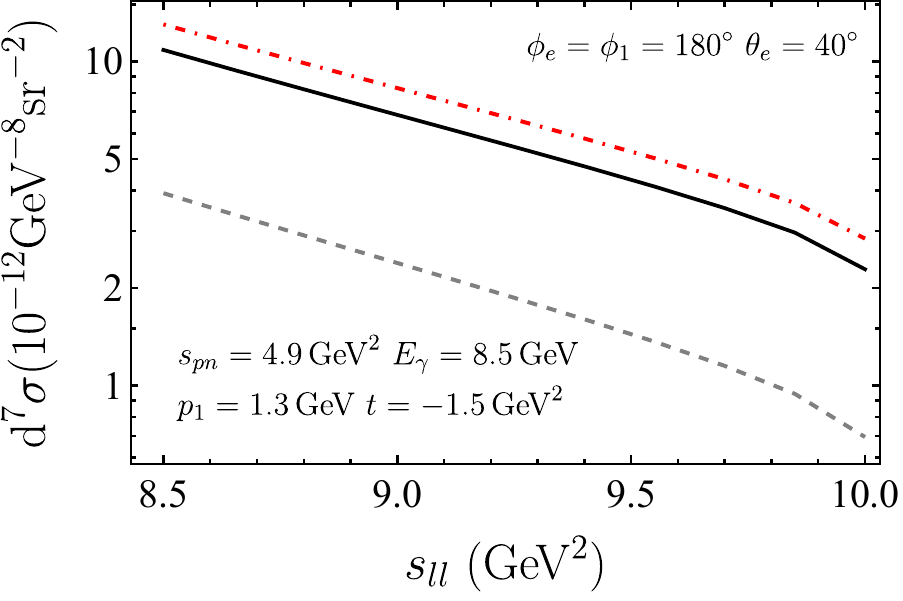}
    \caption{Differential cross sections for $e^+e^-$ production as a function of $s_{ll}$ at $E_\gamma=8.5~\rm{GeV}$, $s_{pn}=4.9~\rm{GeV}^2$, $t=-1.5~\rm{GeV}^2$, $p_1=1.3~\rm{GeV}$, $\theta_e=40^\circ$, and $\phi_1=\phi_e=180^\circ$. The gray dashed curve represents the result with PWIA, the black solid curve represents the result including the on-shell FSI, and the red dot-dashed curve represents the result including both the on-shell and the off-shell FSIs.}
    \label{fig_fsi_sll}
\end{figure}

To clearly see the contribution from the PWIA and FSI interference term and that from the FSI amplitude square term separately, we plot the differential cross section as a function of $s_{pn}$ with other variables fixed at $t=-1.7~\rm{GeV}^2$, $s_{ll}=9~\rm{GeV}^2$, $\phi_1=180^\circ$, $\theta_e=60^\circ$, and $\phi_e=90^\circ$ in Fig.~\ref{fig_fsi_spn}, where the dashed curve represent the result with PWIA, the dot-dashed curve includes the interference term between the PWIA and FSI amplitudes, and the solid curve includes both the interference term and the FSI amplitude square term. As can be observed, the interference term generally suppress the cross section.
For a chosen missing momentum, {\it e.g.} $p_m=0.45~\rm{GeV}$, there exists a relation between the polar angle of the missing momentum with respect to the virtual photon, $\theta_m$, and the $s_{pn}$, as marked on the top and bottom axes in Fig.~\ref{fig_fsi_spn}. At forward angles of the missing momentum, the interference term dominants FSI effects and the cross section is suppressed in comparison with the PWIA result even if both the interference term and the FSI amplitude square term are included. However, when the $\theta_m$ is large, which is more likely to be caused by the NN rescattering effects, the differential cross section is enhanced by the FSI and reaches a peak around $\theta_m \sim 65^\circ$ for the particular kinematic choice in Fig.~\ref{fig_fsi_spn}. This is consistent with the typical FSI effects as mentioned in Refs.~\cite{Frankfurt:1996xx,Sargsian:2001ax,Laget:2004sm,Jeschonnek:2008zg} that the recoil nucleon receives transverse momentum through the rescattering while the knocked-out nucleon tends to eject away from the forward direction. 

In Fig.~\ref{fig_fsi_sll}, we plot the cross section as a function of $s_{ll}$ with other variables fixed at $s_{pn}=4.9~\rm{GeV}^2$, $t=-1.5~\rm{GeV}^2$, $p_1=1.3~\rm{GeV}$, $\theta_e=40^\circ$, and $\phi_1=\phi_e=180^\circ$. As has been discussed in Sec.~\ref{singular_behavior}, at large $\theta_e$, {\it e.g.} $\theta_e=40^\circ$, the lepton momentum is away from the photon collinear direction and thus there is not a sharp peak in the $s_{ll}$ dependence for this particular kinematic choice. The FSI enhances the differential cross section but does not distort the trend of the curve in this case.

\subsection{Interference between the proton and the neutron}

\begin{figure}[htp]
\includegraphics[width=0.45\textwidth]{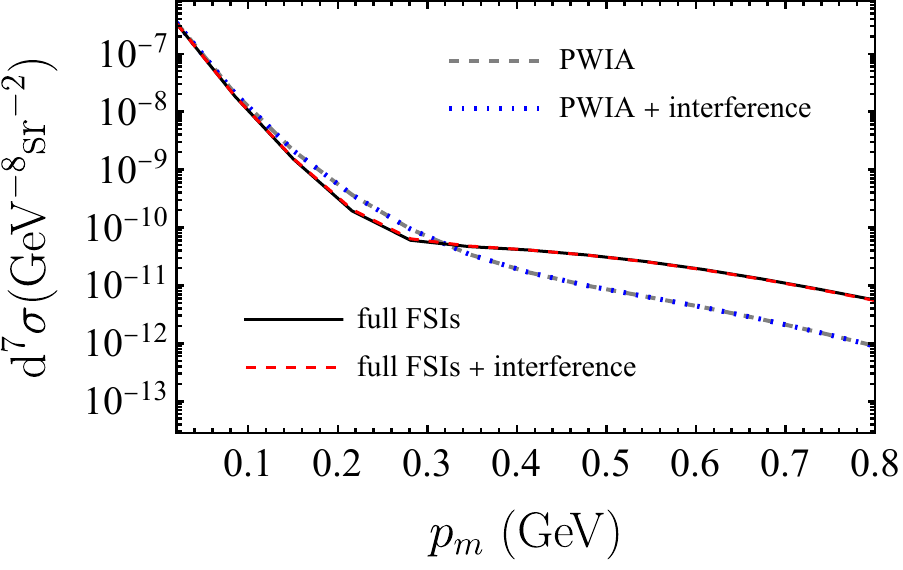}
\caption{The differential cross sections with and without the interference term between the proton and the neutron amplitudes. The results are drawn at $E_\gamma=8.5~\rm{GeV}$, $s_{pn}=5.5~\rm{GeV}^2$, $t=-2~\rm{GeV}^2$, $s_{ll}=m_{J/\psi}^2$, $\theta_e=5^\circ$, and $\phi_e=\phi_1=180^\circ$. }
\label{fig_interfer}
\end{figure}

For the reaction of deuteron disintegration, as shown in Fig.~\ref{feyn_posi_nega}, the virtual photon can either couple to the proton or couple to the neutron. In principle, one need to add the two possibilities at the amplitude level and then there is an interference term between the proton amplitude and the neutron amplitude. However, such contribution is highly suppressed at certain kinematics as considered in this study. 

In the PWIA, the nucleon struck by the virtual photon acquires large momentum, while the spectator nucleon inherits the fermi motion momentum which is unlikely to be greater than several hundred MeV. If in the final state a high momentum nucleon is detected, supposed to be the proton, it is likely the one coupled to the virtual photon. Once the FSI is taken into account, one may imagine that the detected high momentum nucleon is not directly coupled to the virtual photon but receives the large momentum from the NN scattering. In this case we need to demonstrate whether the interference contribution is still negligible. Assuming the isospin symmetry, the electromagnetic current operator matrix element is given by
\begin{equation}
    J^\mu(q)_{\rm{if}}=\frac{1}{\sqrt{2}}\left(J_p^\mu(q)_{\rm{if}}-J_n^\mu(q)_{\rm{if}}\right) ~,
\end{equation}
where the subscript $N = p,n$ indicates the nucleon which absorbs the virtual photon. Then the interference term for the nuclear tensor is 
\begin{equation}
    W_{\rm{interf}}^{\mu\nu}=-(2m)^2\overline{\sum_{\rm{if}}}\frac{1}{2}\left(J_p^\mu(q)_{\rm{if}} J_n^{\nu\,\dagger}(q)_{\rm{if}}+J_n^\mu(q)_{\rm{if}} J_p^{\nu\,\dagger}(q)_{\rm{if}}\right)~.
\end{equation}

In Fig.~\ref{fig_interfer}, we compare the differential cross section with and without the interference term between the proton amplitude and the neutron amplitude in the range of the missing momentum from $0$ to $0.8\,\rm GeV$. The gray dashed and blue dotted curves represent the results with and without the interference term in the PWIA, respectively. The black solid and the red dashed curves represent the results with and without the interference term including FSI effects, respectively. No visible difference is observed in either case.

\section{Summary and conclusions}
\label{conclusions}

In this paper, we study the lepton pair production in the photodisintegration of the deuteron. We perform a calculation of the complete seven-fold differential cross section with the lepton pair produced through the Bethe-Heitler mechanism. The deuteron bound state is described by relativistic covariant deuteron-nucleon vertex, in which the form factors are given by linear combinations of deuteron wave functions that are the solution of the Gross equation~\cite{Gross:2010qm}. Apart from the calculation with PWIA, we take into account the FSI, which arises from the rescattering between the knock-out nucleon and the spectator nucleon. The NN scattering amplitude is parameterized in a general Lorentz covariant form including all spin-dependent contributions, and the parametrization functions are derived from the partial wave expansion of the Saclay formalism.

We also carry out numerical calculations. The differential cross section can vary by orders of magnitude in dependence on the lepton azimuthal angle $\phi_l$ when the polar angle $\theta_l$ is approaching $0^\circ$ or $180^\circ$, while for the middle polar angle the azimuthal angle dependence is mild. In the $s_{ll}$ and $s_{pn}$ dependence, sharp peaks are also observed. All these nearly singular behaviors arise from the collinear singularity for massless leptons when the lepton or the antilepton is moving along the incidence photon direction. At the certain kinematics investigated in this study, the produced lepton pair is moving in forward angle, and thus the sharp peak does not appear when the lepton polar angle is near vertical angle. This singularity is physically regularized by keeping the lepton mass in our calculation and numerically we observe that the differential cross section for $\mu^+\mu^-$ production is much smoother than that for $e^+e^-$ due to the heavier mass.

The FSI effects have strong dependence on the missing momentum $p_m$, because the recoil nucleon can receive energy-momentum transfer in the rescattering. When the missing momentum is relatively small, the FSI contribution is negligible and the PWIA provides a good approximation to the full result. While the missing momentum is large, the FSI leads to significant enhancement of the differential cross section. In the intermediate missing momentum region, we observe a suppression of the cross section due to the interference between the PWIA amplitude and the FSI amplitude. Separating the FSI contribution into the on-shell part and the off-shell part, we find that the on-shell term dominates at relatively small $|t|$ region in which the Bjorken $x$ goes close to 1, the quasi-elastic limit. When $x$ is away from $1$, corresponding to large $|t|$ region, the off-shell term dominates the FSI effect. Furthermore, we investigate the contribution from the interference between the proton amplitude and the neutron amplitude. With explicit numerical calculation, we demonstrate that this contribution is negligible in the kinematics region considered in this study, even if the FSI effects are taken into account.

The dilepton production in deuteron photodisintegration process can not only provide us valuable information to understand the deuteron structure but also serve as an important input in the analysis of $J/\psi$ production, deeply virtual meson production, and other exclusive measurements using deuteron target or beam. Therefore, it is a very relevant process for the physics program at JLab and future electron-ion colliders.

\begin{acknowledgements}
We are grateful to Ron L. Workman and Igor I. Strakovsky for sending us the SAID manual that contains information on the nucleon-nucleon amplitudes. We thank Haiyan Gao, Gregory Matousek, Richard Tyson, and Zhiwen Zhao for helpful discussions. This work was supported by the National Natural Science Foundation of China under contract No. 12075003 and No. 12175117.
\end{acknowledgements}

\appendix
\section{Differential cross section \label{dcs_formula}}
The four body phase space is
 \begin{equation}
     \int \mathrm{d} \Pi= \int (2\pi)^{4} \delta^{(4)}(k+P-p_1-p_2-p_3-p_4)\frac{\mathrm{d}^3 \bm{p}_1}{(2\pi)^3 2 E_1}  \frac{\mathrm{d}^3 \bm{p}_2}{(2\pi)^3 2 E_2}  \frac{\mathrm{d}^3 \bm{p}_3}{(2\pi)^3 2 E_3} \frac{\mathrm{d}^3 \bm{p}_4}{(2\pi)^3 2 E_4}~.
 \end{equation}
 To separate the phase space for the lepton pair part and the nuclear part, the following delta function integration is inserted,
 \begin{eqnarray}
     \int \frac{\mathrm{d}^4 p_{34}}{(2\pi)^4} \,(2\pi)^4 \,\delta^{(4)}(p_{34}-p_3-p_4) \int \frac{\mathrm{d}s_{ll}}{2\pi} \,2\pi\, \delta(s_{ll}-p_{34}^2)\nonumber\\
     =\int\frac{\mathrm{d}s_{ll}}{2\pi}\int \frac{\mathrm{d}^3 \bm{p}_{34}}{(2\pi)^3 2E_{34}}\,(2\pi)^4\, \delta^{(4)}(p_{34}-p_3-p_4)~.
 \end{eqnarray}
 Where $p^0_{34}=E_{34}=\sqrt{|\bm{p}_{34}|^2+s_{ll}} $. Then, the phase space is 
 \begin{eqnarray}
     \int \mathrm{d} \Pi &=& \int\frac{\mathrm{d}s_{ll}}{2\pi}  (2\pi)^4 \delta^{(4)}(k+P-p_1-p_2-p_{34}) \frac{\mathrm{d}^3 \bm{p}_1}{(2\pi)^3 2 E_1}  \frac{\mathrm{d}^3 \bm{p}_2}{(2\pi)^3 2 E_2} \frac{\mathrm{d}^3 \bm{p}_{34}}{(2\pi)^3 2E_{34}} \nonumber\\
     &\times& (2\pi)^4 \,\delta^{(4)}(p_{34}-p_3-p_4) \frac{\mathrm{d}^3 \bm{p}_3}{(2\pi)^3 2 E_3}  \frac{\mathrm{d}^3 \bm{p}_4}{(2\pi)^3 2 E_4}~.
 \end{eqnarray}
 The phase space for lepton pair is calculated in the lepton pair center of mass frame,
 \begin{eqnarray}
      \mathrm{d}\Pi_{ll}&=&\int (2\pi)^4 \,\delta^{(4)}(p_{34}-p_3-p_4) \frac{\mathrm{d}^3 \bm{p}_3}{(2\pi)^3 2 E_3}  \frac{\mathrm{d}^3 \bm{p}_4}{(2\pi)^3 2 E_4}\nonumber\\
     &=&\int \frac{|\bm{p}_c|}{\sqrt{s_{ll}}}\frac{\mathrm{d} \Omega_{ll}}{(4\pi)^2}~,
 \end{eqnarray}
 in which $\bm{p}_c$ is the momenta of the lepton in the center of mass frame. The phase space for the proton and the neutron is calculated in the laboratory system, in which $z$ axis is along $\bm{q}$ direction,
 \begin{eqnarray}
     \mathrm{d} \Pi_{pn}&=&\int (2\pi)^4 \delta^{(4)}(k+P-p_1-p_2-p_{34}) \frac{\mathrm{d}^3 \bm{p}_1}{(2\pi)^3 2 E_1}  \frac{\mathrm{d}^3 \bm{p}_2}{(2\pi)^3 2 E_2} \frac{\mathrm{d}^3 \bm{p}_{34}}{(2\pi)^3 2E_{34}}\nonumber\\
     &=&\int \frac{1}{(2\pi)^5}\frac{|\bm{p}_1|^2}{8E_{34}}\frac{\mathrm{d}^3 |\bm{p}_{34}|\mathrm{d} \Omega_1}{||\bm{p}_1|(m_d+\nu)-E_1|\bm{q}|\cos \theta_1|}~.
     \label{psproneu}
 \end{eqnarray}
 where $\bm{p}_1$ is determined by the equation 
 \begin{equation}
     m_d+\nu=\sqrt{m^2+|\bm{p}_1|^2}+\sqrt{m^2+|\bm{p}_1|^2+|\bm{q}|^2-2|\bm{q}||\bm{p}_1|\cos \theta_1}~.
 \end{equation}
 Using the following relations,
 \begin{eqnarray}
     s_{pn}&=&t+2m_d\nu+m_d^2~,\nonumber\\
     |\bm{p}_{34}|&=&\sqrt{(E_\gamma-\nu)^2-s_{ll}}~,\nonumber\\
     \cos\theta_{34}&=&\frac{E_\gamma^2+(E_\gamma-\nu)^2-s_{ll}-(\nu^2-t)}{2E_\gamma\sqrt{(E_\gamma-\nu)^2-s_{ll}}}~,
 \end{eqnarray}
one can make variables transformation from $s_{ll}$, $|\bm{p}_{34}|$, and $\theta_{34}$ to $s_{ll}$, $s_{pn}$, and $t$. With Eqs.~(\ref{dcs_equation}) and (\ref{ampsq}), the differential cross section is expressed by
 \begin{eqnarray}
     \diff \sigma =\frac{\alpha^3}{8(4\pi)^4}\frac{|\bm{p}_1|^2\beta}{m_d^2 E_\gamma^2 t^2}\frac{L^{\mu\nu}W_{\mu\nu}}{||\bm{p}_1|(m_d+\nu)-E_1|\bm{q}|\cos \theta_1|}{\rm{d}} \Omega_p {\rm{d}} \Omega_{ll}{\rm{d}}s_{ll}{\rm{d}}s_{pn}{\rm{d}}t~.
 \end{eqnarray}
\section{Relations between scalar invariant functions and deuteron wave functions \label{rela_form_wave}}
Realtions between four scalar invariant functions and deuteron wave functions in the deuteron rest frame are given by
\begin{eqnarray}
    F(\bm{p})&=&\pi \sqrt{2 m_{d}}\left(2 E_{p}-m_{d}\right)\left[u(\bm{p})-\frac{1}{\sqrt{2}} w(\bm{p})+\sqrt{\frac{3}{2}}\frac{m}{|\bm{p}|} v_{t}(\bm{p})\right]~, \\
	G(\bm{p})&=&\pi m\sqrt{2 m_{d}}\left(2 E_{p}-m_{d}\right)\left[\frac{ u(\bm{p})}{E_{p}+m}+ \frac{\left(2 E_{p}+m\right)}{\sqrt{2}|\bm{p}|^{2}}w(\bm{p})+\sqrt{\frac{3}{2}}\frac{1}{|\bm{p}|} v_{t}(\bm{p})\right]~,\\
	H(\bm{p})&=&\pi \sqrt{3m_{d}} \frac{E_{p} m}{|\bm{p}|} v_{t}(\bm{p})~, \\
	I(\bm{p})&=&-\pi  \frac{\sqrt{2}m^{2}}{\sqrt{m_{d}}}\left[\left(2 E_{p}-m_{d}\right)\left(\frac{u(\bm{p})}{E_{p}+m}-  \frac{E_{p}+2 m}{\sqrt{2}|\bm{p}|^2}w(\bm{p})\right)+\sqrt{3}\frac{m_{d}}{|\bm{p}|}  v_{s}(\bm{p})\right]~,
\end{eqnarray}
where $|\bm{p}|=\sqrt{\dfrac{(P\cdot k_2)^2}{P^2}-k_2^2}$, $P$ is the four-momentum of the deuteron and $k_2$ is the four-momentum of the on-shell nucleon. The derivation details  can be found in Refs.~\cite{Buck:1979ff,Gross:2014wqa}. The normalization of wave functions is 
\begin{equation}
    \int_0^\infty \left[u^2(\bm{p})+w^2(\bm{p})+v_t^2(\bm{p})+v_s^2(\bm{p})\right] |\bm{p}|^2 \diff |\bm{p}| = 1~.
\end{equation}
In our calculation, numerical solutions for the deuteron wave functions $u(\bm{p})$, $w(\bm{p})$, $v_t(\bm{p})$ and $v_s(\bm{p})$ are from Ref.~\cite{Gross:2010qm}.
\section{Plane wave contribution to the nuclear tensor \label{trace_formu}}
we show construction details for the Dirac trace in Eq.~(\ref{pw_trace}) in this appendix. With Eq.~(\ref{pw_contri}), $W^{\mu\nu}$ is written in the form 
\begin{eqnarray}
    W^{\mu\nu}&=&\frac{(2m)^2}{3}\sum_{\lambda_d,s_1,s_2}\left\langle\bm{p}_{1} s_{1} ; \bm{p}_{2} s_{2}\left|\hat{J}^{\mu}\right| \bm{P} \lambda_{d}\right\rangle_{\rm{PW}}\left\langle \bm{P} \lambda_{d}\left|\hat{J}^{\nu\,\dagger}\right|\bm{p}_{1} s_{1} ; \bm{p}_{2} s_{2}\right\rangle_{\rm{PW}}\nonumber\\
    &=&\frac{(2m)^2}{3}\sum_{\lambda_d,s_1,s_2}\bar{u}\left(\bm{p}_{1}, s_{1}\right) \Gamma_N^{\mu}(q) \frac{\slashed{P}-\slashed{p}_2+m}{m^2-(P-p_2)^2} \Gamma_{\lambda_{d}}\left(p_{2}, P\right)C \gamma^0 u^{*} \left(\bm{p}_{2}, s_{2}\right)\nonumber\\
	&\times& \bar{u}^{*} \left(\bm{p}_{2}, s_{2}\right)C^\dagger \gamma^0 \tilde{\Gamma}_{\lambda_d}\frac{\slashed{P}-\slashed{p}_2+m}{m^2-(P-p_2)^2}\tilde{\Gamma}_N^\nu u\left(\bm{p}_{1}, s_{1}\right)~.
\end{eqnarray}
Note that $\gamma^0 \gamma^{\mu\,\dagger} \gamma^0=\gamma^\mu$, $\gamma^0 \gamma^{\mu\,*} \gamma^0=\gamma^{\mu\,T}$, $\gamma^0 C^\dagger \gamma^0 = -C^{-1}$, and $C \gamma^\mu C^{-1}= - \gamma^{\mu\,T}$,
\begin{eqnarray}
    \sum_{s_2} C \gamma^0 u^{*} \left(\bm{p}_{2}, s_{2}\right) \bar{u}^{*} \left(\bm{p}_{2}, s_{2}\right)C^\dagger \gamma^0=\frac{\slashed{p}_2-m}{2m}~.
\end{eqnarray}
It then follows that
\begin{eqnarray}
    W^{\mu\nu} &=&{\rm{Tr}}\left[\Gamma_N^{\mu}(q) \frac{\slashed{P}-\slashed{p}_2+m}{m^2-(P-p_2)^2} \Gamma_D^\alpha\left(p_{2}, P\right)\left(\slashed{p}_2-m\right)\widetilde{\Gamma}_D^\beta\frac{\slashed{P}-\slashed{p}_2+m}{m^2-(P-p_2)^2}\widetilde{\Gamma}_N^\nu \left(\slashed{p}_1+m\right)\right]\nonumber\\
	&\times& \frac{1}{3}\sum_{\lambda_d=1}^3 \xi_\alpha^{\lambda_d}(P)~\xi^{*\,\lambda_d}_\beta(P)~.
\end{eqnarray}
\section{Parametrization for NN elastic scattering matrix \label{para_M}}
Two ways of parameterization in Eqs.~(\ref{cov_para}) and (\ref{pauli_para}) for the nucleon nucleon scattering matrix can be related by the helicity amplitudes. Helicity amplitudes are defined by scattering amplitudes in which initial states and final states have the specific helicity $\lambda_i$ and $\lambda^\prime_i$,
\begin{equation}
\mathcal{M}_{\lambda_{1}^{\prime} \lambda_{2}^{\prime} ; \lambda_{1} \lambda_{2}} =\bar{u}_{\alpha^\prime}\left(\bm{p}_{1}^{\prime},\lambda_{1}^{\prime}\right)\bar{u}_{\beta^\prime}\left(\bm{p}_{2}^{\prime},\lambda_{2}^{\prime}\right) M_{\alpha^\prime \alpha, \beta^\prime \beta}\,u_{\alpha}\left(\bm{p}_{1},\lambda_{1}\right)u_{\beta}\left(\bm{p}_{2},\lambda_{2}\right)~.
\end{equation}
where $\alpha$, $\alpha^\prime$ and $\beta$, $\beta^\prime$  are the Dirac index for the first and the second particle, respectively. $u\left(\bm{p}_{i},\lambda_{i}\right)$ is the Dirac spinor with momentum $\bm{p}_i$ and helicity $\lambda_i$ and is defined by
\begin{equation}
    u(\bm{p}_i, \lambda)=N\begin{pmatrix}
	1 \\
	2 \lambda \tilde{p}_i
	\end{pmatrix} | \lambda\rangle_i~,
	\label{four_spinor}
\end{equation}
where $\tilde{p}=\dfrac{|\bm{p}|}{E_p+m}$, $N=\sqrt{\dfrac{E_p+m}{2m}}$ and $E_p=\sqrt{m^2+\bm{p}^2}$.  $|\lambda\rangle_i$ in the center of mass frame is listed in Table \ref{two_spinor}. We use the convention in Refs.~\cite{Jacob:1959at,Gross:1991pm,Gross:2010qm}.

\begin{table*}[h]
    \caption{$|\lambda\rangle_i$ in Eq.~(\ref{four_spinor}) }
    \label{two_spinor}
    \begin{ruledtabular}
    \begin{tabular}{c c c | c c c}
    & Initial state  & & & Final state  & \\
    &    $\lambda=\frac{1}{2}$  & $\lambda=-\frac{1}{2}$&  & $\lambda=\frac{1}{2}$  & $\lambda=-\frac{1}{2}$\\
    \hline
    $|\lambda\rangle_1$& $\begin{pmatrix} 1\\0\end{pmatrix}$&$\begin{pmatrix} 0\\1\end{pmatrix}$&$|\lambda\rangle_1$& $\begin{pmatrix} \cos \frac{1}{2} \theta \\ \sin \frac{1}{2} \theta \end{pmatrix}$&$\begin{pmatrix} -\sin \frac{1}{2} \theta\\ \cos \frac{1}{2} \theta\end{pmatrix}$\\
    
    $|\lambda\rangle_2$& $\begin{pmatrix} 0\\1\end{pmatrix}$&$\begin{pmatrix} 1\\0\end{pmatrix}$&$|\lambda\rangle_2$& $\begin{pmatrix} -\sin \frac{1}{2} \theta\\\cos \frac{1}{2} \theta\end{pmatrix}$&$\begin{pmatrix} \cos \frac{1}{2} \theta\\\sin \frac{1}{2} \theta\end{pmatrix}$\\
    \end{tabular}
    \end{ruledtabular}
\end{table*}
Considering parity conservation, time reversal invariance and the Pauli principle, there are five independent helicity amplitudes \cite{Bystricky:1976jr},
\begin{eqnarray}
    M_{1} &\equiv& \mathcal{M}_{\frac{1}{2}\frac{1}{2};\frac{1}{2}\frac{1}{2}}~,\nonumber\\
    M_{2} &\equiv& \mathcal{M}_{\frac{1}{2}\frac{1}{2};-\frac{1}{2}-\frac{1}{2}}~, \nonumber\\
	M_{3} &\equiv& \mathcal{M}_{\frac{1}{2}-\frac{1}{2};\frac{1}{2}-\frac{1}{2}}~,\nonumber\\
	M_{4} &\equiv& \mathcal{M}_{\frac{1}{2}-\frac{1}{2};-\frac{1}{2}\frac{1}{2}}~, \nonumber\\
	M_{5} &\equiv& \mathcal{M}_{\frac{1}{2}\frac{1}{2};\frac{1}{2}-\frac{1}{2}}~.
	\label{indep_heli}
\end{eqnarray}
Substituting Dirac spinors with specific helicities from Eq.~(\ref{four_spinor})  and the parameterization Eq.~(\ref{cov_para}) of $M_{\alpha^\prime \alpha, \beta^\prime \beta}$ into Eq.~(\ref{indep_heli}) to calculate independent helicity amplitudes, we obtain the relations between $M_1$, $M_2$, $M_3$, $M_4$, $M_5$ and $F_{S}(s, t)$, $F_{V}(s, t)$, $F_{T}(s, t)$, $F_{P}(s, t)$, $F_{A}(s, t)$,
\begin{eqnarray}
    M_1&=&\frac{1}{2m^2}\left(m^2(\cos \theta+1)F_{S}+ (m^2+m^2\cos \theta +4 \bm{p}^2)F_{V}\right.\nonumber\\
	&+ & \left. 2m^2(\cos \theta -3) F_{T}+(-m^2\cos \theta+3m^2+4\bm{p}^2)F_{A}\right)~,\nonumber\\
	M_2&=&\frac{1}{2m^2}\left((\cos\theta -1)E_p^2F_{S}+ m^2(\cos\theta-1)F_{V}\right.\nonumber\\
	&+&\left.2(\cos \theta +3)(m^2+2\bm{p}^2)F_{T}+ \bm{p}^2(\cos\theta-1)F_{P} -m^2(\cos \theta+3)F_{A}\right)~,\nonumber\\
	M_3&=& \frac{1}{m^2}\left(m^2\cos^2 \frac{\theta}{2}F_{S} +(m^2+2\bm{p}^2)\cos^2\frac{\theta}{2}F_{V}\right.\nonumber\\
    &+&\left.2m^2\cos^2 \frac{\theta}{2}F_{T}-(m^2+2\bm{p}^2)\cos^2 \frac{\theta}{2}F_{A}\right)~,\nonumber\\
	M_4&=&\frac{1}{m^2}\left(\sin^2\frac{\theta}{2}E_p^2F_{S}+m^2\sin^2\frac{\theta}{2}F_{V}\right.\nonumber\\
	&+&\left.2m^2\sin^2\frac{\theta}{2}F_{T}-\bm{p}^2\sin^2\frac{\theta}{2}F_{P}-m^2\sin^2\frac{\theta}{2}F_{A}\right)~,\nonumber\\
	M_5&=&\frac{E_p}{2m}\left(-\sin\theta F_{S} -\sin\theta F_{V}
	-2\sin\theta F_{T}+ \sin\theta F_{A}\right)~.
\end{eqnarray}
Relations between $a$, $b$, $c$, $d$, $e$ and helicity amplitudes $M_i^{\rm SAID}$ are given in Ref.~\cite{Bystricky:1976jr} and listed here for convenience,
\begin{eqnarray}
    M^{\rm SAID}_{1}&=&\frac{1}{2}(a \cos \theta+b-c+d+i e \sin \theta)~, \nonumber\\
	M^{\rm SAID}_{2}&=&\frac{1}{2}(a \cos \theta-b+c+d+i e \sin \theta)~, \nonumber \\
	M^{\rm SAID}_{3}&=&\frac{1}{2}(a \cos \theta+b+c-d+i e \sin \theta)~, \nonumber \\
	M^{\rm SAID}_{4}&=&\frac{1}{2}(-a \cos \theta+b+c+d-i e \sin \theta)~, \nonumber \\
	M^{\rm SAID}_{5}&=&\frac{1}{2}(-a \sin \theta+i e \cos \theta)~.
\end{eqnarray}
The partial wave expansion for amplitudes $a$, $b$, $c$, $d$ and $e$ is discussed in Ref.~\cite{Bystricky1987}. Using the phase shift analysis of Refs.~\cite{Arndt:1982ep,Workman:2016ysf} and phase shift data available from SAID program~\cite{said_program}, we obtain the value of amplitudes at the specific momentum $|\bm{p}|$ and scattering angle $\theta$. The unpolarized differential cross section in our convention is given by
 \begin{equation}
     \frac{\diff \sigma}{\diff \Omega} = \frac{1}{4} \sum_{\lambda_1 \lambda_1^\prime \lambda_2 \lambda_2^\prime } |\mathcal{M}_{\lambda_{1}^{\prime} \lambda_{2}^{\prime} ; \lambda_{1} \lambda_{2}}|^2 \frac{(2m)^4}{(2\pi)^2}\frac{1}{16s}~.
 \end{equation}
 In SAID program's convention,
 \begin{equation}
     \frac{\diff \sigma}{\diff \Omega} = \frac{1}{4} \sum_{\lambda_1 \lambda_1^\prime \lambda_2 \lambda_2^\prime } |\mathcal{M}^{\rm SAID}_{\lambda_{1}^{\prime} \lambda_{2}^{\prime} ; \lambda_{1} \lambda_{2}}|^2~.
 \end{equation}
 An additional factor is needed,
 \begin{equation}
     M_i=\frac{2\pi\sqrt{s}}{m^2}M^{\rm SAID}_i~.
 \end{equation}

%

\end{document}